\documentclass[11pt]{article}
\usepackage{amsfonts}
\usepackage{amsmath}
\usepackage{amssymb}
\usepackage{graphicx}
\usepackage{tikz}
\usepackage{amsthm}
\makeatletter
\def\th@plain{%
  \thm@notefont{}%
  \itshape
}
\def\th@definition{%
  \thm@notefont{}%
  \normalfont
}
\makeatother
\usepackage{array}
\usepackage{indentfirst}
\usepackage[left=1in,top=1in,bottom=1in,right=1in]{geometry}
\usepackage[round]{natbib}
\usepackage{verbatim}
\usepackage{enumerate}
\usepackage[export]{adjustbox}
\usepackage{float}
\usepackage{graphics,color}
\usepackage{mathpazo}
\usepackage{xr-hyper}
\usepackage[nodisplayskipstretch]{setspace}
\usepackage[%
plainpages,%
linkcolor=blue,
colorlinks=true,
urlcolor=blue,%
filecolor=black,%
citecolor=black,%
pdfpagemode=UseOutlines,%
pdfauthor={Cole Wittbrodt},%
pdfsubject={Optimal Platform Design},%
]{hyperref}
\usepackage{authblk}

\newtheoremstyle{pfof}
{\topsep}
{\topsep}
{\normalfont}
{0pt}
{\bfseries}
{}
{5pt plus 1pt minus 1pt}
{}
\theoremstyle{pfof}
\newtheorem*{proofof}{Proof of}
\theoremstyle{definition}
\newtheorem*{defn}{Definition}
\newtheorem*{assumption}{Assumption}

\newtheorem*{rem}{Remark}

\theoremstyle{plain}
\newtheorem{thm}{Theorem}

\newtheorem{prop}{Proposition}
\newtheorem{cor}{Corollary}
\newtheorem{lem}{Lemma}

\theoremstyle{remark}

\DeclareMathOperator{\supp}{supp}
\newcommand{\hi}{h}
\newcommand{\lo}{\ell}

\title{Optimal Platforms for Search and Matching}
\author{Cole Wittbrodt\thanks{Wittbrodt: Columbia University, Graduate School of Arts and Sciences, Department of Economics; E-mail: \texttt{cole.wittbrodt@columbia.edu}. This paper was previously circulated under the title 'Optimal Platform Design' and studied a more limited environment. I thank Antoine Chapel, Yeon-Koo Che, Grace Chuan, Laura Doval, Navin Kartik, Qingmin Liu, Alessandro Pavan, Zhihui Wang, Nathan Yoder, Dylan Yost, and conference participants at the Econometric Society World Congress for helpful comments and feedback.}}

\begin{document}

\maketitle

\begin{abstract}
Search and matching increasingly take place on online platforms with elements of centralized and decentralized matching. Platforms can alter the search process for users, but cannot eliminate search frictions entirely. I study a model where users pay to change the \textit{distribution} of potential partners they search over and characterize the revenue-maximizing platform when match characteristics are private information. With complementary match characteristics, I show that the only possible source of inefficiency on the optimal platform is exclusion, unlike standard screening problems. Matching on the optimal platform is perfectly assortative --- there is no equilibrium mismatch. This conclusion is robust to various manners in which fees are collected (e.g. subscription fees or match fees), and to environments where the platform designer can also change search speeds.
\end{abstract}

\newpage
\section{Introduction}

In many settings, economic agents spend a considerable amount of time searching for a suitable match; firms and workers exert tremendous effort to fill jobs in labor markets, tourists search for the ideal vacation rental, and individuals conduct prolonged searches for a romantic partner. Increasingly, these searches take place on online platforms, such as Linkedin, Airbnb, or Hinge. However, these platforms cannot eliminate search frictions entirely. Users may not always be available to match, matches can vary in duration, and a platform cannot generally force two users who meet to form a match. Instead, platforms affect the search process by changing the distribution of agents one searches over. Real world search on platforms, thus, has elements of both centralized and decentralized matching.

In this paper, I study a strategic search and matching process on such platforms with a focus on optimal design for revenue maximizing platforms. I model a dynamic environment where searching for a partner is time intensive and agents possess heterogeneous match characteristics. The platform designer does not know the match characteristics of agents at the time of contracting. The platform modifies the search and matching procedure by altering the distribution over other agents that one may meet on the platform. In practice, platforms can alter these search distributions through recommendations, rankings, filters, and other forms of targeting, which change the composition of agents a user encounters without necessarily changing the frequency with which search opportunities arrive.\footnote{This paper isolates the effect of changing the search distribution. I also consider an extension in which the platform can change the frequency at which search opportunities arise.} For instance, Linkedin may design their platform in such a way that firms that report a high productivity meet workers that report a high productivity more frequently. %

When there are two agent types (a low-productivity type and a high-productivity type) and match characteristics are complementary, I characterize the set of equilibria of the matching game induced by a particular platform. Under a simple condition ensuring that low types are willing to match with one another (which I refer to as the no-holdout condition), equilibria exist and take a relatively simple form. I show that steady state output, and hence total welfare, is decreasing in the probability of mismatch.\footnote{I demonstrate in Section \ref{S:holdout} that this need not be the case when the no-holdout condition fails.} My main result is that the platform designer's revenue is \textit{also} decreasing in the probability of mismatch, even though match characteristics are private. Hence, the revenue maximizing platform is efficient and assortative conditional on inclusion.

The main force behind this result is that the information rent (i.e. the net payoff of a high type that reports low) is \textit{increasing} in the probability of mismatch. When the probability of mismatch increases, high types have a higher probability of meeting other high types after a deviation. This only encourages high types to misreport their type, given any subsequent equilibrium of the search and matching game. Therefore, the standard screening logic that distortions of the low type's allocation reduce the incentive for high types to deviate fails. Since distortions in the search distributions reduce surplus and increase rent, the designer optimally chooses the assortative platform. The only possible source of inefficiency on the optimal platform is thus the exclusion of low types.

Concretely, I work in a model with an infinitely lived population of agents who must form bilateral matches in order to produce. For simplicity, the matching market is one-sided.\footnote{The analysis that follows can be easily extended to a two-sided matching market.} Agents have one of two match characteristics, $x\in\{\lo,\hi\}$, which are privately observed by the agents but not by the platform designer. Searching for potential partners is time intensive, existing matches dissolve exogenously, and agents can reject a potential partner after observing the partner's type. When two agents match, they produce output according to a strictly supermodular production function, so match characteristics are complements. Utility is transferable and agents bargain over the net surplus generated by a match.

Platforms must also satisfy a market clearing condition: every meeting involves two agents. Market clearing requires the same mass of cross-type search on both sides of the market. Conditional on serving both types, every platform can therefore be represented by a single number $\epsilon$, representing the probability of mismatch. The case $\epsilon=0$ is perfectly assortative; any $\epsilon>0$ involves mismatch.

Before considering the question of optimal platform design, I characterize the equilibrium set under the no-holdout condition. My solution concept is similar to that of \cite{shimersmith2000} by requiring the unmatched fraction of each type of agent to be in steady state. The set of equilibria naturally depends on the platform under consideration. When the probability of mismatch is high, I show that the equilibrium is \textit{accepting}, meaning that all types are willing to match with one another. When the mismatch probability is low, another type of equilibrium can emerge: a \textit{selective equilibrium}. In a selective equilibrium, types are only willing to match with other agents of the same type.    

A monopolist, or \textit{platform designer}, offers a menu of search distribution/fee pairs.\footnote{I show in Appendix \ref{S:DRP} that the restriction to menu mechanisms is without loss of generality. That is, the designer does not benefit from randomization.} In the baseline model, fees are flow payments made to the platform at any point in time. I later consider extensions of the model where payments are collected only while unmatched, only while matched, or only at the time of a match. Motivated by applications such as Linkedin or Hinge, a user can change their choice from a menu at any point in time while unmatched. When an agent reports $\hat{x}$, they draw the \textit{report} of some other agent $\hat{x}'$ from the distribution $g_{\hat{x}}\in\Delta(\{\lo,\hi\})$. When a platform is incentive compatible, $g_{\hat{x}}$ is the true distribution of agent types met after reporting $\hat{x}$. I show that the complicated dynamic incentive constraints can be reduced to a simpler set of incentive constraints that require no type to have a profitable \textit{persistent} deviation. Moreover, the local downward incentive constraints must bind for the revenue maximizing platform.

When the platform designer seeks to maximize their own profits and the no-holdout condition is satisfied, I show that the optimal platform is \textit{conditionally efficient} and assortative. Conditional on search, assortativity maximizes both steady state output and the designer's revenues. A searching agent only ever meets other agent with match characteristics identical to their own. 

The design of an optimal platform is a non-standard screening problem. For contrast, consider a standard second degree price discrimination problem where a monopolist sells an object with endogenous quality (e.g. \cite{mussarosen1978}). There, reducing the quality allocated to a low type makes the low type's contract less attractive to a high type, reducing informational rent. One might expect a platform to do the same thing by distorting the low type's search distribution. This intuition is wrong in the present setting. If match characteristics are complements, high types have a comparative advantage in matching with high types. Putting more high types in the low search distribution therefore makes the low type's allocation \textit{more} attractive to a high type (under the no-holdout condition). A distortion thus reduces the value generated by matching at the same time that it increases the informational rent of the high type.

Matching behavior resembles that of \cite{becker1973}. There is no equilibrium mismatch. On the optimal platform, the sole role of search frictions is that agents will spend some time unmatched. Moreover, agents do not spend an inefficiently long time unmatched; every meeting on the optimal platform results in a match. This differs substantially from the frictional benchmark without a platform designer. In the absence of a platform designer, there would either be mismatch or agents would waste time meeting others with whom they would be unwilling to match.

The dynamic nature of the matching market is important for understanding the scope of this result. When low types are willing to match with one another, search frictions generate delay but do not create a motive for the platform to distort the composition of search. When low types instead reject one another, introducing mismatch can change the steady-state population of agents available to search. In particular, keeping low types unmatched for longer can make them more available to high types in future meetings. This availability channel has no counterpart in a static assignment problem and can make positive mismatch either privately or socially desirable.

I consider a number of extensions to the baseline model, and show that assortativity is robust. I first consider variations on the timing of payments. Surprisingly, when fees are collected only while agents are unmatched, a revenue maximizing designer has no incentive to keep agents unmatched for longer periods by distorting the platform. I also consider an extension where the platform designer can alter not only the search distribution of users, but also their meeting rates. I show that the only source of distortion on the intensive margin is an inefficiently low meeting rate for the low type if the equilibrium associated with the assortative platform is selective. Standard screening once again applies; by reducing the low type's meeting rate the designer reduces the high type's incentive to purchase the low type's menu object.

\subsection*{Related Literature}

This article is closely related to work which studies the design of matching intermediaries. \cite{damianoli2007} and \cite{johnson2013}, for instance, study the design of one-to-one matching intermediaries, whereas \cite{gomespavan2016} study a many-to-many intermediary. More recently, \cite{aoyagiyoo2022} study a two-sided platform which chooses matching probabilities and subscription fees when matched agents subsequently interact strategically. These papers consider settings where an intermediary either chooses matches directly or sorts agents into sets of possible matches. The present article instead studies a strategic dynamic search environment. The platform changes the distribution of agents that a user searches over, but agents must still be simultaneously available and a meeting forms a match only if both parties agree. Search frictions therefore make the population of agents available to match endogenous to the platform. This distinction is important for results: assortativity is seldom optimal in this literature.\footnote{\cite{damianoli2007} is the exception, and discussed in more detail in Section \ref{S:discuss}.}

The paper is also related to work which studies platform design in dynamic search markets. \cite{niedermayershneyerov2014} study optimal pricing by a profit-maximizing intermediary in a dynamic two-sided search market with private information. Their focus is on the pricing and participation decisions of the intermediary, whereas the present article studies how a platform optimally designs the composition of search. \cite{romanyuksmolin2019} instead allow a platform to design the information observed by agents before they decide whether to accept a potential match. In the present article, match characteristics are observed at the time of a meeting, and the platform changes which types of agents meet rather than the information revealed about a given meeting. \cite{immorlicaetal2023} studies an environment where a platform guides decentralized search by controlling meeting opportunities or market accessibility. Their objective is the design of approximately efficient or welfare-maximizing search, whereas I study a revenue-maximizing platform when match characteristics are private information. \cite{tehwangwatanabe2024} studies a similar environment for a profit maximizing intermediary, but without ex-ante heterogeneity and private user types.

The paper is also related to models of directed and partially directed search. In these environments, agents direct their search toward different submarkets or posted contracts, and equilibrium sorting can cause agents to search only over particular types of trading partners; see \cite{wrightetal2021} for a survey. Other papers model improvements in search technology through the frequency or direction of meetings rather than through their composition. For example, \cite{farboodijaroschshimer2023} allow agents to invest in their contact rates, while \cite{wu2024} studies costly directed search. These instruments are distinct from the search distributions studied here. A meeting rate determines how frequently an agent receives opportunities, whereas a search distribution determines the composition of those opportunities. This distinction is consequential for screening. As I show in Section \ref{S:intensity}, reducing the meeting rate offered to a low type can reduce the information rent of a high type and therefore generates the standard screening tradeoff. Changing the composition of search does not generate this standard screening tradeoff.

Finally, the paper is related to the decentralized search and matching literature. Classic contributions include \cite{diamond1982}, \cite{mortensen1982}, and \cite{pissarides2000}. Much of this literature studies conditions under which matching is assortative. \cite{shimersmith2000} studies transferable utility in a frictional search environment and extends the assortative-matching logic of \cite{becker1973}. Relative to this literature, I allow a platform to change the distribution of meetings and charge agents for access to different search distributions. Importantly, the platform cannot dictate matches or eliminate the underlying availability constraint. The interaction between this market-clearing constraint and private information is what distinguishes the screening problem studied here from standard models of price discrimination.

\section{Matching Model}\label{S:ExMark}

There is a unit mass of infinitely lived agents. Time is continuous. Each agent has a type $x\in X:=\{\lo,\hi\}$. The probability that an agent is a high type $(\hi)$ is $\pi\in(0,1)$, so high types have population mass $\pi$ and low types $(\lo)$ have population mass $1-\pi$. Type $\hi$ is interpreted as having a more productive or desirable match characteristic than type $\lo$.

The total output when types $x,y$ match is $f_{xy}>0$. Assume that $f_{\lo\hi}=f_{\hi\lo}$. I assume that output is strictly increasing in each argument, $f_{\lo\lo}<f_{\lo\hi}<f_{\hi\hi}$, and strictly supermodular,
\begin{align}\label{E:SPM}\tag{SPM}
    f_{\lo\lo}+f_{\hi\hi}>2f_{\lo\hi}.
\end{align}
Thus, match characteristics are complements. Holding fixed the number of matches, replacing two low-high matches with a low-low match and a high-high match raises total output.

\subsubsection*{Platforms and Market Clearing}

Agents choose whether or not to engage with a \textit{platform}. A platform specifies a search distribution for each reported type, and each search distribution has an associated price. Formally, a platform is a stationary menu of the form
\begin{align*}
    P=\{(g_{\lo},t_{\lo}),(g_{\hi},t_{\hi})\},
\end{align*}
where $g_x\in\Delta(\{\lo,\hi\})\cup\{\emptyset\}$ is the search distribution designed for type $x$ and $t_x\in\mathbb R$ is a flow payment.\footnote{Stationary is a natural assumption in the baseline model, as I study \textbf{steady-state equilibria} of the ensuing matching game for tractability. The restriction to menus is without loss of generality, as the designer does not benefit from stochastic stationary mechanisms (see Appendix \ref{S:DRP}).} I use $g_x=\emptyset$ to denote exclusion. An excluded agent is unable to search and cannot be met by an included agent, so an excluded agent receives a gross payoff of zero. Throughout, let $\tilde{X}$ denote the set of included types. In the baseline model, an included agent pays $t_x$ at each moment in time, whether matched or unmatched. I consider alternative payment schemes in Section \ref{S:alt}.

Platforms must satisfy a \textit{market clearing} condition. Market clearing simply ensures that meetings are bilateral. When both types are included, the mass of low types directed toward high types must equal the mass of high types directed toward low types:
\begin{align}\label{E:MCC}\tag{MCC}
    (1-\pi)g_{\lo}(\hi)=\pi g_{\hi}(\lo).
\end{align}
Let $\epsilon$ denote this common mass of cross-type search. Every full-inclusion platform can therefore be written as
\begin{align*}
    g_{\lo}^{\epsilon}
    &=
    \left(1-\frac{\epsilon}{1-\pi}\right)\delta_{\lo}
    +\frac{\epsilon}{1-\pi}\delta_{\hi},\\
    g_{\hi}^{\epsilon}
    &=
    \frac{\epsilon}{\pi}\delta_{\lo}
    +\left(1-\frac{\epsilon}{\pi}\right)\delta_{\hi},
\end{align*}
where $0\leq\epsilon\leq\bar{\epsilon}:=\min\{\pi,1-\pi\}$.\footnote{Throughout, I use $\delta_x$ to denote the Dirac measure on $x\in\{\lo,\hi\}$.} I refer to $\epsilon$ as the \textit{mismatch rate}. A value of $\epsilon=0$ corresponds to a platform where each agent meets only their own type (which I call an assortative platform). A value of $\epsilon=\pi(1-\pi)$ corresponds to the case where $g_x$ is the population distribution for each $x$ (which corresponds to the case of \cite{shimersmith2000}).

\subsubsection*{Search and Matching}

Once on the platform, each included agent searches for a potential match partner. This process closely follows the canonical frictional search framework in \cite{shimersmith2000}, with the following key distinction: the search process is governed by the platform. An agent is either matched or unmatched at any point in time. When two unmatched agents meet, they observe each other's type and can send a "match request". If each agent requests to match, a match is formed and each agent enters the pool of matched agents. Let $u_x\in[0,1]$ denote the fraction of unmatched agents of type $x$.\footnote{That is, the total mass of unmatched agents is $\pi u_\hi+(1-\pi)u_\lo$.} Agents are "called to meet" at a flow rate of $\rho>0$.\footnote{That is, meetings occur according to a continuous time Poisson process with rate parameter $\rho$.} Any matched agent that is called to meet "misses their meeting" and is unable to form a match.\footnote{In the literature on search and matching, this is often referred to as \textit{quadratic} search technology. \cite{shimersmith2000} assume quadratic search technology for tractability. Importantly, under quadratic search technology the payoffs of an agent of type $x$ are unaffected by the strategy of any agent unwilling to match with $x$. \cite{lauermannetal} show that steady state equilibria exist under a much richer class of search technologies.} Searching is otherwise costless. Matches dissolve at an exogenous divorce rate of $\alpha>0$.\footnote{That is, divorces occur according to a continuous time Poisson process with rate parameter $\alpha$.} Agents discount the future at a rate $r>0$.

I assume that utility is transferable.\footnote{Qualitatively, the model extends easily to the case where utility is non-transferable. I present the baseline model with transferable utility to match the job search application.} When agents form a match, they bargain over the surplus according to Nash bargaining. In the Nash bargaining solution, agents split the match surplus (the match output $f_{xy}$ net of each agent's expected flow utility from remaining unmatched) evenly.

A strategy for an agent of type $x$ is a set of agents (included or excluded) with whom they would be willing to match. Denote this set $\mathcal{A}(x)$. If $y\in\mathcal{A}(x)$ and $x\in\mathcal{A}(y)$, then when unmatched agents $x$ and $y$ meet, they form a match. This allows one to define a matching sets $\mathcal{M}(x)$, which is the sets of mutually acceptable matches for type $x$. Formally, $\mathcal{M}(x)=\{y:y\in\mathcal{A}(x)\;\text{and}\;x\in\mathcal{A}(y)\}$.

Given a platform $P$, a directed steady state search equilibrium is a profile of strategies for agents $\mathcal{A}$ and fractions of unmatched agents $u$ such that everyone chooses a set of agents to match with that would maximize their expected payoff and the population of unmatched agents is in steady state --- provided an agent of type $x$ chooses menu item $(g_x,t_x)\in P$. Moreover, I impose the following counterfactual acceptance condition: if $x$ finds matching with $y$ to be profitable, but $y\not\in\supp g_x$ (i.e. $x$ can never meet $y$), then $y\in\mathcal{A}(x)$. That is, \textit{all} acceptable matches to an agent (regardless of whether or not they can be met on-path) are included in an agent's acceptance set.

I characterize directed steady state search equilibrium with three types of objects: reservation wages $w_\lo$ and $w_\hi$ that denotes the expected flow match utility for agents (conditional on being unmatched) as a function of their match characteristics, matching sets $\mathcal{M}(\lo)$ and $\mathcal{M}(\hi)$, and steady-state unmatched fractions $u_\lo$ and $u_\hi$.

When agents $x$ and $y$ match, the net surplus is $f_{xy}-w_x-w_y$, since each agent's outside option is their reservation wage. According to the Nash bargaining solution, the agents will split this value after discounting and accounting for the exogenous divorce probability. This allows one, given agent strategies $\mathcal{A}$, to implicitly characterize the reservation wage for included types $x\in\tilde{X}$ as 
\begin{align*}
    w_x=\theta\sum_{y\in\mathcal M(x)}\left(f_{xy}-w_x-w_y\right)u_y g_x(y),
\end{align*}
where I define $\theta=\frac{\rho}{2(r+\alpha)}$. For $x\not\in\tilde{X}$, I naturally define $w_x=0$.

In a directed steady-state equilibrium, the agents' strategies must satisfy the following optimality condition: a match is acceptable whenever the net flow payoffs from matching are positive.\footnote{I break ties by assuming there is no match when agents are indifferent (i.e. when $f_{xy}=w_x+w_y$).} That is, 
\begin{align*}
    \mathcal{A}(x)=\{y:f_{xy}-w_x-w_y>0\}.
\end{align*}
If $y\in\mathcal{A}(x)$, then $x\in\mathcal{A}(y)$ by symmetry of the production function $f$. So $\mathcal{A}(x)=\mathcal{M}(x)$ in a directed steady state equilibrium. 

Included agents of type $x\in\tilde{X}$ enter the market due to divorce, so the inflow mass of type $x$ agents is $\alpha(1-u_x)$. Agents of type $x$ exit the market when they form a match. So, the unmatched density of agents of type $x\in\tilde{X}$ must satisfy the following balance condition in equilibrium
\begin{align*}
    \alpha(1-u_x)
    =\rho u_x\sum_{y\in\mathcal M(x)}u_y g_x(y)
\end{align*}
All excluded agents are unmatched, so $u_x=1$ for all $x\not\in\tilde{X}$. The following definition summarizes these equilibrium conditions.

\begin{defn}[Directed Steady State Search Equilibrium]
      Given a platform $P$, a \textit{directed steady state search equilibrium} (DSE) is a triplet $(w,\mathcal{M},u)$ consisting of reservation wages $w=(w_\lo,w_\hi)$, matching sets $\mathcal{M}=(\mathcal{M}(\lo),\mathcal{M}(\hi))$, and unmatched fractions $u=(u_\lo,u_\hi)$ such that

    \begin{enumerate}[(i)]
        \item Given $\mathcal{M}$, $w_x$ satisfies the Bellman equation for all $x\in\tilde{X}$. That is,
        \begin{align*}
        w_x=\theta\sum_{y\in\mathcal M(x)}\left(f_{xy}-w_x-w_y\right)u_y g_x(y).
\end{align*}
        \item Given $w$, $\mathcal{M}(x)$ is optimal for all $x$. That is,
        \begin{align*}
            f_{xy}-w_x-w_y>0\iff y\in\mathcal{M}(x).
        \end{align*}
        \item Given $\mathcal{M}$, the economy is in steady state. That is, for all $x\in\tilde{X}$
        \begin{align*}
    \alpha(1-u_x)
    =\rho u_x\sum_{y\in\mathcal M(x)}u_y g_x(y)
\end{align*}
        \item If $x\not\in\tilde{X}$, $w_x=0$ and $u_x=1$.
    \end{enumerate}
\end{defn}

\subsubsection*{The Assortative Benchmark}

I now introduce a particularly important platform, and its unique associated DSE. An assortative platform $P=\{(g_\lo^0,t_\lo),(g_\hi^0,t_h)\}$ satisfies $g_\lo^0=\delta_\lo$ and $g_\hi^0=\delta_\hi$. Every agent searches over other agents that share their type. Own-type meetings must be accepted: if type $x$ rejected its only possible partner, its reservation wage would be zero, making the own-type match strictly profitable. That is, $x\in\mathcal{M}(x)$ in any DSE associated with an assortative platform. The unmatched fraction is therefore the same for both types and solves
\begin{align*}
    \alpha(1-u^0)=\rho(u^0)^2.
\end{align*}
Thus,
\begin{align*}
    u^0=\frac{-\alpha+\sqrt{\alpha^2+4\alpha\rho}}{2\rho}.
\end{align*}
The reservation wages on the assortative platform are
\begin{align*}
    w_{\lo}^0
    &=\frac{\theta u^0}{1+2\theta u^0}f_{\lo\lo},\\
    w_{\hi}^0
    &=\frac{\theta u^0}{1+2\theta u^0}f_{\hi\hi}.
\end{align*}

I refer to the reservation wages, matching correspondences, and unmatched fractions defined above as the \textit{assortative DSE}. The next result confirms that this is the unique DSE corresponding to an assortative platform.

\begin{prop}[Assortative DSE]\label{P:assortDSE}
    If $P$ is an assortative platform and $(w,\mathcal{M},u)$ is a DSE given $P$, then for each $x\in\{\lo,\hi\}$, $w_x=w_x^0$, $x\in\mathcal{M}(x)$, and $u_x=u_x^0$.
\end{prop}

\section{Equilibrium and Efficiency}
\subsection{Equilibrium Characterization}

I now characterize the set of equilibria on a particular platform $P$ before turning to the question of platform design. Characterizing the equilibrium set is trivial whenever there is exclusion.\footnote{For instance, if the low-type is excluded then the only $g_h$ satisfying the market clearing condition is $g_h=\delta_h$. It immediately follows that $w_h=w_h^0$ and $u_h=u^0$.} Therefore, I focus on settings where both types are included. The first result characterizes matching sets that can occur in some equilibrium.

\begin{lem}\label{L:regimes}
    Let $P$ be a platform with $\tilde{X}=X$ and let $(w,\mathcal{M},u)$ be an associated DSE. Then exactly one of the three following cases must hold.
    \begin{enumerate}
        \item (Selective Regime): For each $x\in X$, $\mathcal{M}(x)=\{x\}$.
        \item (Accepting Regime) For each $x\in X$, $\mathcal{M}(x)=\{\lo,\hi\}$.
        \item (Holdout Regime) For the low type, $\mathcal{M}(\lo)=\{\hi\}$ and for the high type $\mathcal{M}(\hi)=\{\lo,\hi\}$.
    \end{enumerate}
\end{lem} 

In the selective regime, types are only willing to match with other agents of the same type. In the accepting regime, every meeting results in a match. The holdout regime is asymmetric; low types are unwilling to match with one another, but the high type finds any match acceptable.

For a platform $P$ with mismatch parameter $\epsilon$, let $\mathcal{E}(\epsilon)$ denote the set of DSE. In general, $\mathcal{E}(\epsilon)$ is difficult to characterize. Existence of DSE is not even guaranteed.\footnote{Standard proofs of existence do not apply to this setting, since I restrict attention to pure match acceptance strategies. If agents were allowed to mix when indifferent, i.e. when $f_{xy}=w_x+w_y$, then standard fixed point arguments can be used to establish the existence of a mixed strategy steady state equilibrium.} Nonetheless, I give a simple primitive condition under which DSE exist and the set of DSE is well-behaved. In particular, this condition rules out the holdout regime of Lemma \ref{L:regimes}. It ensures that $\lo\in\mathcal{M}(\lo)$ for \textit{any} platform.

\begin{assumption}[No-Holdout Condition]
The no-holdout condition is satisfied if
\begin{align*}
    f_{\lo\lo}>
    \frac{2\theta u^0}{1+\theta u^0}f_{\lo\hi}.
\end{align*}
\end{assumption}

The no-holdout condition ensures that a low type never finds it worthwhile to reject another low type in order to remain available for a high partner. It requires low-low output to be large relative to the largest reservation wage a low type can obtain by searching for high partners (in particular, when $\epsilon=\bar{\epsilon}=\min\{\pi,1-\pi\}$). When the option value of continued search is small, an agent has little reason to reject an own-type match in the hope of meeting a high type later.

As I will show shortly, there is at most one equilibrium in the selective regime and at most one equilibrium in the accepting regime for each $\epsilon$. Hence, under the no-holdout condition there are at most two equilibria. In the subsequent paragraphs, I characterize the equilibria in each regime.

Fix $\epsilon\in[0,\bar{\epsilon}]$. Suppose that there is a selective DSE. That is, $\mathcal{M}(x)=\{x\}$. Letting $u_x^S(\epsilon)$ satisfy the balance condition given $\mathcal{M}$, we have that
\begin{align*}
    \alpha(1-u_\lo^S(\epsilon))&=\rho\left(1-\frac{\epsilon}{1-\pi}\right)(u_\lo^S(\epsilon))^2\\\alpha(1-u_\hi^S(\epsilon))&=\rho\left(1-\frac{\epsilon}{\pi}\right)(u_\hi^S(\epsilon))^2.
\end{align*}
Hence, $u_x^S(\epsilon)$ is uniquely determined. Additionally,
\begin{align*}
    w_\lo^S(\epsilon)&=\frac{\theta\left(1-\frac{\epsilon}{1-\pi}\right) u_\lo^S(\epsilon)}{1+2\theta\left(1-\frac{\epsilon}{1-\pi}\right) u_\lo^S(\epsilon)}f_{\lo\lo},\\w_\hi^S(\epsilon)&=\frac{\theta\left(1-\frac{\epsilon}{\pi}\right) u_\hi^S(\epsilon)}{1+2\theta\left(1-\frac{\epsilon}{\pi}\right) u_\hi^S(\epsilon)}f_{\hi\hi}
\end{align*}
uniquely solve the Bellman equation for each type. Hence, a selective equilibrium exists if and only if 
\begin{align*}
    f_{\lo\hi}-w_\lo^S(\epsilon)-w_h^S(\epsilon)\leq0.
\end{align*}
The selective equilibrium (when it exists) admits simple comparative statics. In particular, the unmatched fraction of agents is increasing in $\epsilon$ and the reservation wages are decreasing in $\epsilon$. 
\begin{lem}\label{L:mon}
    Suppose that a selective DSE exists for both $\epsilon,\epsilon'\in[0,\bar{\epsilon}]$ with $\epsilon'>\epsilon$. Then $u_x^S(\epsilon')>u_x^S(\epsilon)$ and $w_x^S(\epsilon')<w_x^S(\epsilon)$.
\end{lem} 

It follows from the above lemma that the set of mismatch rates for which a selective equilibrium exists is a lower set, since $f_{\lo\hi}-w_\lo^S(\epsilon)-w_h^S(\epsilon)$ is strictly increasing in $\epsilon$. Hence, $\exists \epsilon^D$ such that a selective equilibrium exists if and only if $\epsilon\leq \epsilon^D$.\footnote{It is possible to have $\epsilon^D<0$, hence a selective equilibrium does not exist for any $\epsilon$.}

Now instead suppose there is an accepting DSE. That is, $\mathcal{M}(x)=\{\lo,\hi\}$. Closed form expressions are more difficult to obtain in this setting, and comparative statics are similarly difficult.\footnote{Although, since every meeting results in a match, $u_x^A(\epsilon)=u^0$ for each $x$.} Nonetheless, there is a unique accepting equilibrium.

\begin{lem}\label{L:uniqueacc}
    If there exists an accepting equilibrium for $\epsilon$, it is unique.
\end{lem}
Similarly to the analysis of the selective regime, $\exists\epsilon^A$ such that an accepting equilibrium exists if and only if $\epsilon>\epsilon^A$.\footnote{It is possible to have $\epsilon^A>\bar{\epsilon}$, hence an accepting equilibrium does not exist for any $\epsilon$.}

I now show that $\epsilon^A\leq\epsilon^D$, with strict inequality except in a knife-edge case. This yields a simple characterization of the set of equilibria. One of three cases must hold: (1) only selective equilibria exist, (2) only accepting equilibria exist, and (3) on some interval, both types of equilibria exist.

\begin{prop}\label{P:regions}
    Suppose the no-holdout condition is satisfied. There exist thresholds $\epsilon^A\leq\epsilon^D$ such that a selective equilibrium exists if and only if $\epsilon\leq\epsilon^D$ and an accepting equilibrium exists if and only if $\epsilon>\epsilon^A$. Moreover, $\epsilon^A<\epsilon^D$ except when $f_{\lo\hi}=w_{\lo}^0+w_{\hi}^0$.
\end{prop}

In particular, under the no-holdout condition, $\mathcal E(\epsilon)\neq\emptyset$ for every feasible mismatch rate $\epsilon\in[0,\bar\epsilon]$.

\begin{figure}[ht]
    \centering
    \begin{tikzpicture}[x=11cm,y=1cm,>=stealth]

        \def\eA{0.38}
        \def\eD{0.68}

        \draw[->] (0,0) -- (1.05,0)
            node[right] {$\epsilon$};

        \draw (0,0.08) -- (0,-0.08)
            node[below=3pt] {$0$};

        \draw (\eA,0.08) -- (\eA,-0.08)
            node[below=3pt] {$\epsilon^A$};

        \draw (\eD,0.08) -- (\eD,-0.08)
            node[below=3pt] {$\epsilon^D$};

        \draw (1,0.08) -- (1,-0.08)
            node[below=3pt] {$\bar\epsilon$};

        \draw[dashed] (\eA,0.15) -- (\eA,2.45);
        \draw[dashed] (\eD,0.15) -- (\eD,2.45);

        \node[left] at (-0.02,2) {Selective};
        \draw[line width=1.5pt] (0,2) -- (\eD,2);
        \fill (0,2) circle (1.7pt);
        \fill (\eD,2) circle (1.7pt);

        \node[left] at (-0.02,1) {Accepting};
        \draw[line width=1.5pt] (\eA,1) -- (1,1);
        \draw[fill=white,line width=.8pt] (\eA,1) circle (1.9pt);
        \fill (1,1) circle (1.7pt);

        \node[align=center] at ({\eA/2},2.75)
            {Selective\\only};

        \node[align=center] at ({(\eA+\eD)/2},2.75)
            {Both DSEs};

        \node[align=center] at ({(\eD+1)/2},2.75)
            {Accepting\\only};

    \end{tikzpicture}
    \caption{Equilibrium regimes under the no-holdout condition.
    The accepting equilibrium emerges before the selective equilibrium
    disappears, so $\epsilon^A<\epsilon^D$ and both equilibria coexist
    for intermediate mismatch rates.}
    \label{F:equilibrium-regimes}
\end{figure}
\subsection{Efficient Platforms}

I consider now the platform that would be chosen by a social planner who can change the search environment, but faces the same matching technology as the platform designer. The planner cannot force two agents to match after they meet. Instead, it chooses the mismatch rate $\epsilon$ and chooses the DSE which follows. Transfers only redistribute surplus, so they play no role in the planner's problem.

The planner maximizes the utility of an agent sampled from the stationary population, averaged over whether they are matched or unmatched. Equivalently, the planner maximizes the flow of output produced by the steady-state stock of matches. Consider a full-inclusion platform with mismatch rate $\epsilon$. The flow rate at which low-low matches form is
\begin{align*}
    \frac{\rho}{2}(1-\pi-\epsilon)u_{\lo}^2
\end{align*}
when low-low meetings are accepted. The factor of one half appears because each low-low match removes two low types from the unmatched pool. Similarly, the flow rate of low-high match formation is $\rho\epsilon u_{\lo}u_{\hi}$, and the flow rate of high-high match formation is $\rho(\pi-\epsilon)u_{\hi}^2/2$. Since matches dissolve at rate $\alpha$, steady-state output is
\begin{align*}
    \mathcal W(\epsilon,w,\mathcal M,u)
    =\frac{\rho}{2\alpha}\Big[&
    (1-\pi-\epsilon)u_{\lo}^2
    \mathbf 1\{\lo\in\mathcal M(\lo)\}f_{\lo\lo}\\
    &+2\epsilon u_{\lo}u_{\hi}
    \mathbf 1\{\hi\in\mathcal M(\lo)\}f_{\lo\hi}\\
    &+(\pi-\epsilon)u_{\hi}^2
    \mathbf 1\{\hi\in\mathcal M(\hi)\}f_{\hi\hi}\Big].
\end{align*}
The social planner solves
\begin{align}\label{E:SPP}\tag{SPP}
    \max_{\epsilon,w,\mathcal M,u}\quad
    &\mathcal W(\epsilon,w,\mathcal M,u)\\
    \text{s.t.}\quad
    &0\leq\epsilon\leq\min\{\pi,1-\pi\}\nonumber,\\
    &(w,\mathcal M,u)\text{ is a DSE induced by }g^\epsilon.\nonumber
\end{align}
It is without loss to require full inclusion in this problem. If one type were excluded, the planner could add that type assortatively without changing the search problem of the included type, generating strictly positive additional output. I call any platform that solves \eqref{E:SPP} \textit{efficient}.

\begin{prop}[Efficient Search]\label{P:efficiency}
Fix a platform $P$ with a DSE $(w,\mathcal{M},u)$. If $\lo\in\mathcal{M}(\lo)$, then
\begin{align*}
    \mathcal W(\epsilon,w,\mathcal M,u)
    \leq
    \mathcal W(0,w^0,\mathcal M^0,u^0),
\end{align*}
with strict inequality whenever $\epsilon>0$.
\end{prop}

The proposition follows from two simple observations. If every meeting is accepted, then $u_{\lo}=u_{\hi}=u^0$ and
\begin{align*}
    \mathcal W(\epsilon)
    =\frac{\rho(u^0)^2}{2\alpha}
    \left[
    (1-\pi)f_{\lo\lo}+\pi f_{\hi\hi}
    -\epsilon(f_{\lo\lo}+f_{\hi\hi}-2f_{\lo\hi})
    \right].
\end{align*}
Thus, supermodularity makes output strictly decreasing in mismatch. If low-high meetings are rejected instead, positive mismatch only sends agents toward partners they refuse. Each type then spends more time unmatched than under assortative search, while all realized matches remain own-type matches. Output again falls.

\begin{cor}\label{C:efficientsearch}
    Suppose the no-holdout condition is satisfied. Then the assortative platform is the unique efficient platform.
\end{cor}

\section{Platform Design}\label{S:analysis}

\subsection{Implementability}
\label{S:dynamicIC}

The designer does not observe agents' types. Every agent selects an element from the menu $P$. If an agent of type $x$ selects $(g_{\hat{x}},t_{\hat{x}})$, he searches according to $g_{\hat x}$ whenever unmatched. In mechanism design language, I say an agent of type $x$ \textit{reports} his type to be $\hat{x}$. At each moment in time, he pays fee $t_{\hat{x}}$. At any point in time an agent is unmatched, they are allowed to change their choice from $P$, or exit the platform.\footnote{This is effectively a restriction on the space of contracts that the designer may offer. It is particularly realistic in applications. Nonetheless, I consider variations on the space of contracts in Section \ref{S:alt}.} The report currently in force determines the agent's search distribution and flow payment. If a match is formed, the report and payment in force at the time of formation remain fixed for the duration of the match. When the match dissolves, the agent is again unmatched and may exit or choose either report.

Appealing to the revelation principle, I study platforms (and their associated steady states) where agents have an incentive to report truthfully at any moment in time.\footnote{See Appendix \ref{S:DRP} for a formalization of this revelation principle.} These incentive constraints are complex. Nonetheless, I show that these incentive constraints are equivalent to a simple steady state condition.

I first state incentive compatibility directly in terms of this dynamic reporting problem. Fix a truthful steady state $(w,\mathcal M,u)$ and a platform $P$. Consider a true type $x$ who is currently unmatched, and let $v\in\mathbb{R}_+$ denote his net flow-equivalent continuation value from returning to the unmatched state. If he currently chooses report $\hat x$, define
\begin{align*}
    \Phi_{x,\hat x}(v)
    =
    -t_{\hat x}
    +
    \theta
    \sum_{y\in\{\lo,\hi\}}
    u_y g_{\hat x}(y)
    \left[
        f_{xy}-v-t_{\hat x}-w_y
    \right]_+.
\end{align*}
To understand this expression, suppose that the agent meets a type $y$.
If the meeting is rejected, type $x$ returns to the unmatched state and
obtains net continuation value $v$. Since the flow payment
$t_{\hat x}$ remains in force if the match is accepted, his corresponding
gross disagreement value in the bargaining problem is
$v+t_{\hat x}$. A truthful type $y$ has net unmatched value
$w_y-t_y$ and pays $t_y$, so his gross disagreement value is $w_y$.
The expression inside the positive part is therefore the flow surplus
generated by the match relative to the two agents' disagreement values.
Symmetric Nash bargaining gives each agent one-half of this surplus.

Let $V_x$ denote the optimal net continuation value of a type $x$ agent
whenever he is unmatched. Since he may exit or choose either report at
any such history, $V_x$ must satisfy the following Bellman equation
\begin{align}\tag{Bell}
    V_x
    =
    \max\left\{
        0,\,
        \Phi_{x,\lo}(V_x),\,
        \Phi_{x,\hi}(V_x)
    \right\}.
    \label{E:dynamicBellman}
\end{align}
The first term corresponds to the ability for the agent to exit the platform, whereas the others correspond to the ability to report either $\hat{x}=\lo$ or $\hat{x}=\hi$. A platform is dynamically individually rational and incentive compatible if truthful reporting attains the maximum in
\eqref{E:dynamicBellman} for each type (i.e. $V_x=\Phi_{x,x}(V_x)$). In a truthful steady state, $\Phi_{x,x}(w_x-t_x)=w_x-t_x$. Hence dynamic individual rationality and incentive compatibility can be written
\begin{align}
    w_x-t_x&\geq0,
    \label{E:dynamicIR}\tag{D-IR}\\
    w_x-t_x
    &\geq
    \Phi_{x,\hat x}(w_x-t_x)
    \qquad\text{for every }\hat x\neq x.
    \label{E:dynamicIC}\tag{D-IC}
\end{align}
The dynamic IR constraint guarantees that, at each moment in time while unmatched, agents prefer to truthfully report $x$ than to exit the platform. The dynamic IC constraint guarantees that the flow payoff from reporting $x$ truthfully exceeds the flow payoff from reporting $\hat{x}$, and then re-optimizing when reaching the unmatched state again.

It will be convenient, however, to consider the value of a persistent deviation. For a true type $x$ and report $\hat x$, let $\tilde w_{x,\hat x}$ solve
\begin{align*}
    \tilde w_{x,\hat x}
    =
    \theta
    \sum_{y\in\{\lo,\hi\}}
    u_y g_{\hat x}(y)
    \left[
        f_{xy}-\tilde w_{x,\hat x}-w_y
    \right]_+.
\end{align*}
Thus $\tilde w_{x,\hat x}$ is the gross reservation wage that type $x$
would obtain if he were to use report $\hat x$ whenever unmatched
forever. In particular, observe that $\tilde w_{x,x}=w_x$. The $[\cdot]_+$ operator in the above expression allows for the following type of deviation to be possible: suppose that the DSE is in the selective regime (so $\mathcal{M}(x)=\{x\}$) and the high type deviates to a report of $\lo$. In the selective regime, low types do not form matches with high types. However, they may find it desirable to form a match with a high type \textit{that has deviated}. In particular, a deviating high type may demand a smaller share of the surplus than a truthful high type, and matches may form even after a deviation.

I now show that it is sufficient to consider only \textit{persistent} misreports in the dynamic incentive problem. This considerably simplifies the analysis of the revenue-maximizing problem studied below.

\begin{prop}[Persistent Deviations Suffice]
\label{P:dynamicIC}
The dynamic individual-rationality and incentive-compatibility conditions
\eqref{E:dynamicIR} and \eqref{E:dynamicIC} hold if and only if
\begin{align}
    w_x-t_x&\geq0,
    \label{E:IR}\tag{IR}\\
    w_x-t_x
    &\geq
    \tilde w_{x,\hat x}-t_{\hat x}
    \qquad\text{for every }\hat x\neq x.
    \label{E:IC}\tag{IC}
\end{align}
\end{prop}

\subsection{Revenue Maximizing Platforms}

I now characterize the revenue-maximizing platform. First, I show that only downward incentive constraints matter, and they must bind for the optimal platform. Second, I show that (contrary to standard screening models) mismatch \textit{raises} the information rent possessed by the high type under the no-holdout condition. Hence a revenue-maximizing platform never uses mismatch. This no distortion at the \textit{bottom} result (conditional on inclusion of the low type) yields a simple characterization of the optimal platform.

Formally, the designer solves the following optimization problem:
\begin{align*}
    REV^*=\sup_{P,(w,\mathcal{M},u)}\pi t_\hi+(1-\pi)t_\lo\qquad\text{s.t.}\begin{cases}
        \eqref{E:dynamicIR}\\\eqref{E:dynamicIC}\\\eqref{E:MCC}\\(w,\mathcal{M},u)\;\text{is a DSE given $P$}.
    \end{cases}
\end{align*}

Proposition \ref{P:dynamicIC} allows one to write the dynamic IR and IC constraints in a simple quasilinear form. One can then apply standard arguments from mechanism design to conclude that only local downward incentive constraints are relevant to a revenue maximizing platform designer.

\begin{lem}\label{L:revenue-screening}
    Consider a pair $P$ and $(w,\mathcal{M},u)$ satisfying \eqref{E:MCC}.
    \begin{enumerate}[(i)]
        \item If $(w,\mathcal{M},u)$ is a DSE given $P$, $t_\lo=w_\lo$, $t_\hi=w_\hi-(\tilde{w}_{\hi,\lo}-w_\lo)$, and a single-crossing condition holds:
        \begin{align}\label{E:SCC}\tag{SCC}
            w_\hi-\tilde{w}_{\hi,\lo}\geq \tilde{w}_{\lo,\hi}-w_\lo
        \end{align}
        then $P$ and $(w,\mathcal{M},u)$ is feasible in the revenue maximization problem.
        \item If $P$ and $(w,\mathcal{M},u)$ is feasible in the revenue maximization problem, it satisfies \eqref{E:SCC}. Moreover, it is revenue-dominated by some platform $\hat{P}$ satisfying $\hat{t}_\lo=w_\lo$ and $\hat{t}_\hi=w_\hi-(\tilde{w}_{\hi,\lo}-w_\lo)$.
    \end{enumerate}
\end{lem}
Hence, the revenue on the optimal platform can be written as
\begin{align*}
    (1-\pi)w_\lo+\pi(w_h-(\tilde{w}_{\hi,\lo}-w_\lo))
\end{align*}
Thus the platform maximizes population-weighted reservation value net of the informational rent that must be left to the mass $\pi$ of high types.

Consider a platform $P$ defined by the mismatch rate $\epsilon$, and consider an associated DSE $(w,\mathcal{M},u)$. If $\mathcal{M}(x)=\{\lo,\hi\}$ (i.e. the DSE is an acceptance equilibrium), the unmatched fractions are then equal to $u^0$ regardless of $\epsilon$. Therefore, changing $\epsilon$ only changes the composition of matches.\footnote{Provided that all matches remain acceptable after changing $\epsilon$ slightly. That is, I consider changes to $\epsilon\in(\epsilon^A,\bar{\epsilon}]$.} Population-weighted reservation wages satisfy
\begin{align*}
    (1-\pi)w_{\lo}+\pi w_{\hi}
    =
    \frac{\theta u^0}{1+2\theta u^0}
    \left[
    (1-\pi)f_{\lo\lo}+\pi f_{\hi\hi}
    -\epsilon(f_{\lo\lo}+f_{\hi\hi}-2f_{\lo\hi})
    \right],
\end{align*}
which is strictly decreasing in $\epsilon$ by supermodularity.

The same conclusion holds if low-high meetings are rejected (i.e. the DSE is a selective equilibrium). In this case, mismatch does not generate different matches at all. It simply directs agents toward partners they will refuse, so each type waits longer for an own-type partner and both reservation wages fall. Hence, whenever low-low meetings remain acceptable, mismatch weakly lowers gross match value.

One might still expect that mismatch helps the platform by reducing informational rent $\tilde{w}_{\hi,\lo}-w_\lo$. It does not. As the low menu becomes less assortative, it contains more high types. Because high types benefit relatively more from high partners, this makes the low menu more attractive to a high type. The high type's rent therefore rises. 

\begin{prop}\label{P:nodist}
Fix an implementable full-inclusion platform $P$ and associated DSE $(w,\mathcal{M},u)$ where $\lo\in\mathcal{M}(\lo)$. Then
\begin{align*}
    (1-\pi)w_{\lo}+\pi w_{\hi}
    &\leq
    (1-\pi)w_{\lo}^0+\pi w_{\hi}^0,\\
    \tilde w_{\hi,\lo}-w_{\lo}
    &\geq
    \frac{\theta u^0}{1+\theta u^0}
    (f_{\lo\hi}-f_{\lo\lo}).
\end{align*}
Both inequalities are strict if $\epsilon>0$. Consequently, every non-assortative full-inclusion platform on which low-low meetings are accepted earns strictly less revenue than assortative full inclusion.
\end{prop}

Proposition \ref{P:nodist} says that, so long as low types are willing to match with one another, distortions in the matching move both parts of the platform's objective in the wrong direction. They reduce the total value the platform can extract and increase the rent it must leave to high types. The usual screening tradeoff (i.e. a distortion reduces surplus \textit{and} reduces rent) does not apply. The only distortion a designer needs to consider is exclusion.

Market clearing is essential for this result. If the platform could independently make the low menu worse while leaving the high menu unchanged, it could potentially reduce the value of imitating a low type.\footnote{I study such a model in Section \ref{S:intensity}.} But a bilateral meeting must appear on both sides of the market. Increasing the mass of high types in the low search distribution simultaneously increases the mass of low types in the high search distribution. The platform therefore cannot simply shift all search distributions toward lower-quality partners, as a monopolist might shift down a quality schedule in a standard screening problem.

Hence, every implementable full-inclusion platform is revenue dominated by the assortative platform under the no-holdout condition. This result is strong and does not depend on whether the DSE considered is in the selective or accepting regime. One can easily verify that the assortative platform satisfies \eqref{E:SCC}, and is thus feasible in the designer's revenue maximization program.\footnote{See Lemma \ref{L:participation} in the Appendix.} This yields a simple characterization of the optimal platform.

\begin{thm}[Optimal Platform]\label{T:secbest}
    Suppose the no holdout condition is satisfied. Then an optimal platform exists. Moreover, on every optimal platform $\epsilon=0$.
\end{thm}

This result differs from the familiar ``no distortion at the top'' logic of standard screening. Here there is no distortion at the bottom either, conditional on serving the low type. If the low type is profitable to serve, he receives the same assortative search distribution that the social planner would choose. If the low menu creates too much rent, the platform excludes the low type. In this region, hidden information affects only the extensive margin of participation.

Theorem \ref{T:secbest} yields a simple characterization of the optimal platform. Only two candidates remain: (i) include both types and choose the assortative platform or (ii) exclude the low type and offer $g_h=\delta_h$ at price $t_h=w_h^0$. 

\begin{cor}\label{C:char}
Suppose the no-holdout condition is satisfied. Then every optimal platform is assortative conditional on participation. More precisely:
\begin{enumerate}[(i)]
    \item If
    \begin{align*}
        \left[1+(1+\pi)\theta u^0\right]f_{\lo\lo}
        >
        \pi(1+2\theta u^0)f_{\lo\hi},
    \end{align*}
    the unique optimal allocation includes both types and sets $g_\lo=\delta_\lo$ and $g_\hi=\delta_\hi$. Revenue-maximizing transfers are $t_\lo=w_\lo^0$ and $t_\hi=w_\hi^0-\frac{\theta u^0}{1+\theta u^0}(f_{\lo\hi}-f_{\lo\lo})$.
    \item If the above displayed inequality is reversed, every optimal platform excludes the low type and sets $g_\hi=\delta_\hi$ and $t_\hi=w_\hi^0$.
    \item If equality holds, both allocations are optimal.
\end{enumerate}
\end{cor}

Equivalently, full inclusion requires
\begin{align*}
    \frac{f_{\lo\lo}}{f_{\lo\hi}}
    \geq
    \frac{\pi(1+2\theta u^0)}
    {1+(1+\pi)\theta u^0}.
\end{align*}
For a fixed $\pi$, the expression on the right increases from $\pi$ toward $2\pi/(1+\pi)$ as $\theta u^0$ rises. Hence, when the option value of continued search is larger, low-low production must be relatively high for the platform to find it worthwhile to introduce the low menu. The cutoff is also increasing in $\pi$. When high types are common, offering a low menu creates informational rent for a larger population; when low types are common, the revenue gained from serving them is correspondingly larger.

When the platform has an incentive to serve low types, the optimal platform is efficient.

\begin{cor}\label{C:optimal-efficient}
    Suppose the no-holdout condition is satisfied and
    \begin{align*}
        \left[1+(1+\pi)\theta u^0\right]f_{\lo\lo}
        >
        \pi(1+2\theta u^0)f_{\lo\hi}.
    \end{align*}
    Then the unique optimal platform solves \eqref{E:SPP}, and is thus efficient.
\end{cor}

\subsection{Discussion}\label{S:discuss}

A handful of modeling assumptions are particularly relevant to the preceding analysis. Most notably, the inability of the platform designer to eliminate search frictions entirely and act as a deterministic matchmaker is important. \cite{damianoli2007} provide a natural static benchmark. As in the present model, two-sided feasibility restricts the quality distortions available to an intermediary. In their environment these restrictions rule out the usual pointwise downward distortion, but leave open coarse matching: pooling neighboring types can raise revenue when the associated virtual match value fails to be sufficiently complementary. In the environment studied here, the conclusion is stronger. Conditional on low types accepting one another, any departure from assortative search both reduces gross match value and raises the informational rent owed to high types. Thus no additional virtual-surplus condition is required for assortative search to be optimal conditional on participation. More fundamentally, the two models differ in what happens after the intermediary chooses the search environment. In Damiano and Li, every participant in a meeting place is randomly assigned a partner with probability one. Here meetings occur over time, agents may reject potential partners, and the distribution of unmatched agents is endogenous. This distinction is particularly important when relaxing the no-holdout condition. When low types are willing to reject one another, mismatch can change who remains available to search and thereby alter future meeting opportunities (as I illustrate in Appendix \ref{A:holdoutexamples}). This availability channel has no counterpart in the static meeting-place model and can make positive mismatch optimal.

The binary-type assumption makes the allocation problem particularly tractable, but the economic force behind Theorem \ref{T:secbest} is not specific to two types. With two types, market clearing reduces every full-inclusion search policy to a single mismatch rate and leaves only one downward incentive constraint. More generally, complementarity implies that higher types have a comparative advantage in higher-quality partners. Thus, shifting a lower type's search distribution toward higher-quality partners makes that allocation relatively more attractive, rather than less attractive, to higher types. At the same time, market clearing prevents the platform from degrading one type's search opportunities in isolation: every change in meeting opportunities must be reflected elsewhere in the market. With richer type spaces, the platform can reshuffle meeting opportunities along many dimensions and must satisfy multiple incentive constraints simultaneously, but the basic idea remains the same. Establishing an analogue of Theorem \ref{T:secbest} with a richer type space requires substantially more notation and analysis. I therefore use the binary environment to more easily convey the economic insights.

Besides conditional efficiency, there are other desirable features of the optimal platform. For each included agent, there is no uncertainty over the match characteristics of the individuals they will meet in the future. In equilibrium, each buyer knows exactly the match characteristics of every agent they meet, but remains uncertain as to \textit{when} those meetings will occur. Additionally, the matching which results from the optimal platform satisfies a certain \textit{conditional stability} property. In particular, within the population of matched agents, there are no blocking pairs.\footnote{That is, there are no pairs of distinct types $x,y\in[0,1]$ such that $x$ prefers matching with $y$ over their equilibrium match partner $x$ and $y$ prefers matching with $x$ over their equilibrium match partner $y$. This follows immediately from \cite{becker1973}.} There are some unmatched agents, however, that could block the equilibrium matching. In standard frictional search and matching models (e.g. \cite{shimersmith2000}), equilibrium matchings seldom satisfy this stability property.\footnote{In a two-type version of \cite{shimersmith2000}, a conditionally stable steady state search equilibrium is only possible if low-high matches are rejected. In such an environment, however, agents on average spend more time unmatched relative to the assortative benchmark.}

\subsection{Beyond the No-Holdout Condition}\label{S:holdout}

When the no-holdout condition is violated, the analysis of platforms becomes substantially more complicated. Equilibrium existence is no longer guaranteed for every $\epsilon\in[0,\bar{\epsilon}]$. More importantly, there are three distinct failures of the analysis in the main model.
\begin{enumerate}
    \item The assortative platform need not be efficient.
    \item The assortative platform need not be optimal.
    \item The optimal platform need not be efficient.
\end{enumerate}
Examples of all three of these instances are given in Appendix \ref{A:holdoutexamples}.

For intuition, suppose that low-high matches are much more productive than low-low matches. Under the assortative platform, every meeting results in a match. By introducing a large probability of mismatch, low types may reject matches with other low types. This raises the unmatched steady state mass of low types, since not every meeting results in a match. High types, therefore, meet more \textit{available} partners. This can increase the proportion of time that high-types spend matched. This can increase the total output of high types in steady state even though $f_{\hi\hi}>f_{\hi\lo}$ since more high types produce at any moment in time. Low type output in steady state also can increase if $f_{\hi\lo}$ is significantly larger than $f_{\lo\lo}$. Hence, even though assortative matching maximizes output in a static assignment problem, it may not maximize output in the dynamic search and matching problem.

The assortative platform need not be optimal for the same reason. Inducing mismatch can now raise their surplus (since they spend more time matched and producing) but has an ambiguous effect on information rents. The impact of introducing a slight probability of mismatch on revenue is thus ambiguous.

Finally, the optimal platform need not be efficient. Moreover, the direction of the distortion is ambiguous. The social planner values mismatch only through its effect on steady-state output, whereas the platform also cares about the informational rent created by the low-type search distribution. In the holdout regime, this rent need not be increasing in mismatch. A truthful low type rejects low partners, while a high type who mimics the low type may still find those partners valuable. Increasing mismatch removes some of these low partners from the low-type search distribution, which can reduce the value of imitation. The platform may therefore introduce more mismatch than the social planner in order to reduce informational rents, even after additional mismatch begins to lower output. Conversely, if mismatch raises informational rents, the platform may introduce less mismatch than the planner even when the resulting availability effect is socially valuable. Thus, there is no general ordering between the revenue-maximizing and efficient levels of mismatch.

\section{Extensions}\label{S:alt}

The baseline model assumes that the platform charges an unconditional flow fee and can change only the composition of search. I now consider two variations. First, I change when the platform collects its fee. Second, I allow the platform to alter the rate at which agents search. Throughout, I maintain the no-holdout condition studied in the baseline model. Both extensions illustrate that the main result is robust under certain conditions.

\subsection{Payment Models}\label{S:payments}

In the baseline model, an agent who reports $x$ makes the flow payment $t_x$ at every moment in time, whether matched or unmatched. Naturally, this payment model is equivalent to one where the designer requests a one-time up front payment.\footnote{This follows from the reduction of \eqref{E:dynamicIR} and \eqref{E:dynamicIC} to the persistent constraints \eqref{E:IR} and \eqref{E:IC}.} This assumption is convenient, since the fee does not affect the match acceptance decision. However, other payment schemes are common in practice. Job search sites and dating apps often charge users while they search, vacation rental platforms collect fees while a match is active, and marriage marketplaces frequently charge a fee when a match is formed. 

A priori, one might expect the assortativity result of the baseline model to fail under alternative payment schemes. For instance, a platform seemingly has an incentive to keep it's users unmatched for longer periods of time when they collect a search fee. This intuition is wrong, however. In this section, I show that the assortative platform is optimal in a larger class of payment schemes. 

I study a more general class of payment timings where the flow payment requested from an agent can depend on their match status but not on the identity of their partner. Moreover, I allow for lump-sum payments to be collected whenever a match forms to. Formally, let $t_x^U$ denote a flow payment made while unmatched, let $t_x^M$ denote a flow payment made while matched, and let $k_x$ denote a lump-sum payment made whenever a match is formed. Four payment models are particularly relevant in applications. Namely, the unconditional flow payment ($t_x^U=t_x^M=t_x$ and $k_x=0$) considered in the baseline model, a pay-while-you-search model ($t_x^M=k_x=0$), a pay-while-matched model ($t_x^U=k_x=0$), and a pay-when-you-match model ($t_x^U=t_x^M=0$).

Unlike the baseline payment, these alternative payment schemes can affect whether a meeting is accepted. A pay-while-you-search fee makes matching more attractive since users may avoid the fee by forming a match. A matched-state fee or match-formation fee has the opposite effect.

Let $U_x$ denote the net flow continuation payoff of a truthful type $x$ while unmatched. Then $w_x=U_x+t_x^U$. Define
\begin{align*}
    \kappa_x
    =
    t_x^M-t_x^U+(r+\alpha)k_x.
\end{align*}
Then the Bellman equation defining reservation wages can be written
\begin{align*}
    w_x
    =
    \theta
    \sum_{y\in\mathcal M(x)}
    u_y g_x(y)
    \left[
        f_{xy}
        -\kappa_x-\kappa_y
        -w_x-w_y
    \right].
\end{align*}
A meeting between $x$ and $y$ is accepted if and only if
\begin{align*}
    f_{xy}
    -\kappa_x-\kappa_y
    -w_x-w_y
    >0.
\end{align*}
The adjusted production function $\widehat{f}_{xy}=f_{xy}-\kappa_x-\kappa_y$ plays the same role as the production function in the baseline model. Importantly,
\begin{align*}
    \widehat f_{\lo\lo}
    +
    \widehat f_{\hi\hi}
    -
    2\widehat f_{\lo\hi}
    =
    f_{\lo\lo}
    +
    f_{\hi\hi}
    -
    2f_{\lo\hi}>0.
\end{align*}
Thus payment timing changes the level of match surplus and can change acceptance behavior, but it does not change the complementarity of match characteristics.\footnote{This would not be true if the designer could implement fees that depend on the type of the matched partner.}

Let $\tilde w_{x,\hat x}$ solve
\begin{align*}
    \tilde w_{x,\hat x}
    =
    \theta
    \sum_{y\in\{\lo,\hi\}}
    u_y g_{\hat x}(y)
    \left[
        f_{xy}
        -\kappa_{\hat x}-\kappa_y
        -\tilde w_{x,\hat x}-w_y
    \right]_+.
\end{align*}
Since an agent using report $\hat x$ has net unmatched utility
$\tilde w_{x,\hat x}-t_{\hat x}^U$, incentive compatibility again takes the form $w_x-t_x^U\geq\tilde{w}_{x,\hat{x}}-t_{\hat{x}}^U$.

Fix the payment rule and associated fees. Let $w_x^0$ and $\tilde w_{\hi,\lo}^0$ denote the values generated by the assortative platform under the same payment terms. Whenever own-type meetings are accepted,
\begin{align*}
    w_x^0
    =
    \frac{\theta u^0}{1+2\theta u^0}
    \left(f_{xx}-2\kappa_x\right).
\end{align*}
Moreover,
\begin{align*}
    \tilde w_{\hi,\lo}^0-w_{\lo}^0
    =
    \frac{\theta u^0}{1+\theta u^0}
    (f_{\lo\hi}-f_{\lo\lo}).
\end{align*}
Notice that the assortative information rent is independent of payment timing. A truthful low type and a high type who chooses the low menu face exactly the same flow payments in expectation.

\begin{prop}[Payment Timing and Mismatch]\label{P:paymenttiming}
Fix a payment scheme, a full-inclusion platform $P$, and an associated DSE $(w,\mathcal M,u)$. Suppose both own-type meetings are accepted: that is $x\in\mathcal{M}(x)$ for $x\in\{\lo,\hi\}$. Then,
\begin{align*}
    (1-\pi)w_{\lo}+\pi w_{\hi}
    &\leq
    (1-\pi)w_{\lo}^0+\pi w_{\hi}^0,\\
    \tilde w_{\hi,\lo}-w_{\lo}
    &\geq
    \tilde w_{\hi,\lo}^0-w_{\lo}^0.
\end{align*}
Both inequalities are strict whenever $\epsilon>0$.

If the adjusted match outputs $\widehat f_{xy}$ is strictly increasing in each argument, then it suffices to assume $\lo\in\mathcal M(\lo)$.
\end{prop}

Proposition \ref{P:paymenttiming} is analogous to Proposition \ref{P:nodist} from the baseline model. Conditional on agents continuing to accept their own types, mismatch has exactly the same two effects as in the baseline model. It lowers the average reservation wage and raises the information rent of the high type from choosing the low menu. Payment timing does not create the standard screening tradeoff.

Intuitively, the complementarity of match characteristics is unchanged (i.e. $\widehat{f}$ is strictly supermodular). Moreover, a high type who chooses the low menu faces the same expected payments as a truthful low type. The high type's relative gain from a low partner remains $f_{\lo\hi}-f_{\lo\lo}$ while its relative gain from a high partner remains $f_{\hi\hi}-f_{\lo\hi}$. The latter is larger by strict supermodularity.

\begin{cor}\label{C:searchfee}
Suppose the no-holdout condition is satisfied and the platform charges nonnegative flow fees only while agents are unmatched. Then every optimal platform is assortative conditional on inclusion.
\end{cor}

This result follows from a simple observation: a fee only paid while searching only makes matches \textit{more} likely to form. Parties are willing to accept a larger set of types, relative to an unconditional fee. Under the no-holdout condition, we have that $x\in\mathcal{M}(x)$ for all types $x$ and for any platform in the baseline payment model. Hence, $x\in\mathcal{M}(x)$ for all types in any equilibrium of any platform with a pay-while-you-search model. Applying Proposition \ref{P:paymenttiming} implies the assortative platform is optimal.

\subsection{Search Intensity}\label{S:intensity}

Theorem \ref{T:secbest} concerns distortions in the composition of search when the rate at which agents receive opportunities is fixed. I now allow the platform to control both margins. I show that, the optimal platform is assortative under a condition on the primitives, but is no longer \textit{conditionally efficient} in general. In particular, the platform designer may distort the meeting rate of the low type to reduce the information rent of the high type. Distortions of this form are unlike the distortions introduced through mismatch. More specifically, on the assortative platform the low type's allocation can be distorted without impacting the allocation of the high type. Hence, standard screening logic obtains.

Formally, a menu is now defined as
\begin{align*}
    P=
    \{(g_\lo,\rho_\lo,t_\lo),(g_\hi,\rho_\hi,t_\hi)\},
\end{align*}
where $\rho_x\in[0,\bar\rho]$ is the rate at which an unmatched agent reporting $x$ is called to meet. I let $\bar{\rho}$ denote an exogenous technological cap on meeting rates. Market clearing requires
\begin{align*}
    (1-\pi)\rho_\lo g_\lo(\hi)
    =
    \pi\rho_\hi g_\hi(\lo).
\end{align*}
I continue to assume unconditional flow payments.

To rule out the availability effect, suppose the no-holdout condition holds uniformly over feasible search rates. A sufficient primitive condition is
\begin{align*}
    f_{\lo\lo}
    >
    \frac{2\bar\rho}{2(r+\alpha)+\bar\rho}
    f_{\lo\hi}.
\end{align*}
Indeed, if a low type rejected another low type, its reservation wage is bounded above by
\begin{align*}
    \frac{\bar\rho}{2(r+\alpha)+\bar\rho}f_{\lo\hi},
\end{align*}
contradicting low-low rejection. Thus, $\lo\in\mathcal{M}(\lo)$ in every DSE.

Endogenous search intensity gives the platform a new way to make the low type's allocation less attractive. The next result shows that, so long as cross-type meetings are rejected, this makes mismatch redundant.

\begin{prop}[Search Intensity]\label{P:searchintensity}
Suppose a full-inclusion platform is associated with a selective DSE. Then it is weakly revenue dominated by an assortative platform. Hence an optimal selective platform is assortative.

Moreover, the optimal selective platform satisfies $\rho_h=\bar{\rho}$ and 
\begin{align*}
    \lambda_\lo:=\rho_\lo u_\lo\in\arg\max_{\lambda\in[\underline\lambda,\bar\rho u^0]}
    \left\{
    (1-\pi)
    \frac{\lambda f_{\lo\lo}}
         {2(r+\alpha+\lambda)}
    -
    \pi
    \frac{\lambda(f_{\lo\hi}-f_{\lo\lo})}
         {2(r+\alpha)+\lambda}
    \right\},
\end{align*}
where $\underline\lambda$ is the smallest matching rate $\rho  u$ that ensures a selective equilibrium exists. 
\end{prop}

Proposition \ref{P:searchintensity} restores the standard screening tradeoff. Slowing the low menu reduces value for low types but also reduces the rent of a high type who imitates them. This differs sharply from a distortion in search composition: within a selective DSE, mismatch merely replaces useful meetings with meetings that are rejected. This distortion reduces surplus for \textit{both} the low and high type, while increasing the rents of the high type. Lowering the meeting rate of the low type reduces surplus available to be extracted from the low type and reduces the information rent possessed by the high type.

However, sufficiently low search intensity can make cross-type meetings attractive. I do not characterize what happens in this case. A simple sufficient condition allows it to be ignored. Let $REV^S$ denote the maximal revenue characterized in Proposition \ref{P:searchintensity} over the set of selective DSE.

\begin{cor}\label{C:selectiveintensity}
If $REV^S>\max\{\pi,1-\pi\}f_{\lo\hi}$, then no optimal full-inclusion platform is associated with an accepting DSE. Consequently, the optimal platform is either the selective platform characterized in Proposition \ref{P:searchintensity} or the high-only platform. If in addition
\begin{align*}
    REV^S\geq \pi w_\hi^0,
\end{align*}
then the optimal platform is the selective full-inclusion platform.
\end{cor}
In the above proposition, $w_\hi^0$ is the reservation wage of high types on assortative platforms with $\rho_h=\bar{\rho}$. 

Search composition is a poor screening instrument because market clearing prevents the platform from degrading the low menu in isolation. Search intensity does not have this feature: the platform can slow the low contract without changing the high contract, and the familiar downward distortion reappears.

\section{Conclusion}

This paper studies the optimal design of matching platforms when the platform designer can alter the distribution of types each agent searches over but cannot otherwise reduce search frictions. I show that the optimal platform is efficient conditional on inclusion under the simple no-holdout condition; if an agent is on the platform, they search over only their own type. The only potential source of inefficiency in this setting is exclusion.

Additionally, on the optimal platform the unique equilibrium unmatched steady state density function depends only on search parameters (the meeting rate and divorce rate). This yields two policy relevant conclusions when applied to a labor market setting. First, in our model, the unmatched fraction is the same for every included type. High skill workers are not less likely to be unemployed than low skill workers, and high productivity firms are not more likely to fill job openings than low productivity firms. All types have the same meeting rate in the matching market and all meetings result in a match on the optimal platform. Second, the unemployment rate does not depend on the payoffs of agents from remaining unmatched.\footnote{Of course, this only holds if the unemployment benefits for type $x$ do not exceed the flow utility of participating in the platform $w^*(x)$.} That is, increasing unemployment benefits does not affect the unemployment rate. This result is consistent with recent empirical findings (e.g. \cite{bdgk}). In a decentralized search and matching market without a platform (as in \cite{shimersmith2000}), increasing unemployment benefits increases the unemployment rate; when remaining unemployed is less costly, agents will opt to remain unmatched for longer to find a higher quality match partner. On the optimal platform, however, unemployment benefits do not distort the agents' incentives, since choosing to remain unemployed for longer cannot result in a better match.

\appendix

\bibliographystyle{ecta}
\bibliography{optimal_platform_design_twotype_no_holdout}

\section{Stationary Mechanisms and the Revelation Principle}
\label{S:DRP}

The baseline model restricts attention to stationary mechanisms. This
restriction is maintained throughout the paper.  Within the stationary class, however, two further restrictions are without loss. First, auxiliary randomization in the platform's meeting rule can be absorbed into a single search distribution. Second, by a revelation argument, an arbitrary stationary mechanism can be replaced by a direct mechanism in which truthful reporting is optimal whenever an agent is unmatched. This section formalizes these reductions.

\subsection{Stationary Mechanisms}

Let $A$ be an arbitrary message space. A \emph{stationary stochastic
mechanism} is a mapping
\begin{align*}
    \Gamma:
    A
    \to
    \Delta(\Delta(X))\times\mathbb R,
\end{align*}
where I denote $\Gamma(a)=(\Gamma_a,t_a)$. Here $t_a\in\mathbb R$ is an expected flow payment and $\Gamma_a\in\Delta(\Delta(X))$ is a stationary stochastic meeting rule associated with message $a$.\footnote{Since agents' preferences are quasilinear in money, treating the flow payments as deterministic is automatically without loss.}

The interpretation is as follows. Whenever an unmatched agent sends
message $a$, he pays the flow fee $t_a$. When he is called to meet, the
platform first draws a search distribution $q\in\Delta(X)$ according to $\Gamma_a$, and then draws a potential partner type
$y\in X$ according to $q$. The draw of $q$ is fresh at every meeting
opportunity and creates no additional persistent state. Conditional on
the current message, therefore, the only payoff-relevant consequence of
this auxiliary randomization is the type of the potential partner whom
the agent meets. A sampled type-$y$ partner is available with probability
$u_y$, as in the search technology in the main text.

A mechanism is \emph{deterministic} if, for every message $a\in A$,
there exists some $g_a\in\Delta(X)$ such that $\Gamma_a=\delta_{g_a}$. In that case I identify the outcome assigned to message $a$ with the pair
$(g_a,t_a)$. The word ``deterministic'' refers only to the mapping from
messages to search distributions and payments. The search distribution
$g_a$ itself is a lottery over potential partner types.

The timing of messages is the same as in the main text. Whenever an agent
is unmatched, he may send any message $a\in A$ or exit the platform. If a
meeting is rejected, he remains unmatched and may immediately send a new
message or exit. If a match is formed, the mechanism outcome associated
with the message in force at the time of formation remains fixed for the
duration of the match. When the match dissolves, the agent again becomes
unmatched and may send any message in $A$ or exit.

An \emph{admissible deviation} for a type $x$ agent is any possibly
history-dependent and randomized rule for choosing an element of
$A\cup\{\emptyset\}$ whenever the agent is unmatched. An agent may change messages repeatedly, condition his choices on his own past meetings and matches, randomize over messages, or exit at any unmatched history.

A stationary implementation of $\Gamma$ consists of a message
$a_x\in A\cup\{\emptyset\}$ for each type $x$ together with a directed
steady-state search equilibrium such that sending $a_x$ whenever
unmatched is sequentially optimal for type $x$ against all admissible
deviations. When $a_x=\emptyset$, type $x$ exits the platform.

For each message $a$, define its reduced-form search distribution by
\begin{align}
    \bar g_a(y)
    =
    \int_{\Delta(X)}
    q(y)\,d\Gamma_a(q),
    \qquad y\in X.
    \label{E:Areducedg}
\end{align}
For a stationary implementation in which both types participate, market
clearing requires
\begin{align}
    (1-\pi)\bar g_{a_{\lo}}(\hi)
    =
    \pi\bar g_{a_{\hi}}(\lo).
    \label{E:AstochasticMC}
\end{align}

\subsection{Randomization in the Meeting Rule}

Two mechanisms are outcome equivalent if under every profile of admissible message and match-acceptance strategies, the two mechanisms generate the same
meeting intensities, continuation payoffs, acceptance decisions, and
steady-state flows. Consequently, they have the same stationary
implementations and generate the same platform revenue. The first result shows that the auxiliary randomization contained in
$\Gamma_a$ is redundant.

\begin{prop}[Stationary Randomization is Redundant]
\label{P:stationary-reduction}
Let $\Gamma:A\to\Delta(\Delta(X))\times \mathbb{R}$ be a stationary stochastic mechanism. Then $\Gamma$ is outcome-equivalent to the
deterministic stationary mechanism $\bar{\Gamma}:A\to\Delta(X)\times\mathbb{R}$ defined by $\bar\Gamma(a)=(\bar g_a,t_a)$.
\end{prop}

\begin{proof}
Fix a message $a\in A$ and a true type $x$. Conditional on sending $a$,
the flow rate at which an unmatched type-$x$ agent meets an unmatched
type-$y$ agent under $\Gamma$ is
\begin{align*}
    \rho u_y
    \int_{\Delta(X)}
    q(y)\,d\Gamma_a(q)
    =
    \rho u_y\bar g_a(y).
\end{align*}
This is exactly the meeting intensity generated by  $\bar g_a$.

More generally, fix arbitrary continuation values and acceptance decisions, and let $S_y$ denote the value to the agent generated by a meeting with type $y$. Linearity of expectations gives
\begin{align*}
    \int_{\Delta(X)}
    \left[
        \sum_{y\in X}
        u_y q(y)S_y
    \right]
    d\Gamma_a(q)
    &=
    \sum_{y\in X}
    u_y
    \left[
        \int_{\Delta(X)}
        q(y)\,d\Gamma_a(q)
    \right]S_y\\
    &=
    \sum_{y\in X}
    u_y\bar g_a(y)S_y.
\end{align*}
Hence every Bellman equation depends on $\Gamma_a$ only through its mean search distribution $\bar g_a$. In particular, under the Nash bargaining
rule used in the main text, replacing $\Gamma_a$ by
$\delta_{\bar g_a}$ leaves the agent's continuation value and match-acceptance decisions unchanged. The same argument applies to the flow out of the unmatched state. For any acceptance rule, the rate at which a type-$x$ agent forms an accepted match after sending $a$ depends only on the meeting intensities $\rho u_y\bar g_a(y)$. Therefore the steady-state balance equations are unchanged. Finally, if $\Gamma$ satisfies market clearing, then
\eqref{E:AstochasticMC} holds. Platform revenue is unchanged because both the equilibrium messages and their associated flow payments are unchanged.
\end{proof}

Proposition \ref{P:stationary-reduction} permits me to restrict attention
to deterministic stationary mechanisms
\begin{align*}
    \Gamma:
    A\longrightarrow\Delta(X)\times\mathbb R,
    \qquad
    \Gamma(a)=(g_a,t_a).
\end{align*}
The next result reduces an arbitrary message space to a direct mechanism.

\subsection{A Stationary Revelation Principle}

Consider an arbitrary deterministic stationary mechanism $\Gamma:A\to\Delta(X)\times\mathbb{R}$ and a stationary implementation in which type $x$ sends message $a_x$ whenever unmatched.

\begin{prop}[Stationary Revelation Principle]
\label{P:stationary-RP}
For every stationary implementation of an arbitrary deterministic
stationary mechanism, there exists a direct stationary mechanism 
\begin{align*}
    \Gamma^D:X\to\left(\Delta(X)\cup\{\emptyset\}\right)\times\mathbb{R}
\end{align*}
which generates the same steady-state allocation and platform revenue and
in which truthful reporting is sequentially optimal at every unmatched
history.
\end{prop}

\begin{proof}
Let $a_x$ denote the message sent by type $x$ in the original stationary
implementation. If type $x$ participates, define
\begin{align*}
    \Gamma^D(x)
    =
    \Gamma(a_x)
    =
    (g_{a_x},t_{a_x}).
\end{align*}
If type $x$ exits in the original implementation, assign $x$ the excluded
outcome.

Under truthful reporting, a type-$x$ agent receives exactly the same
search distribution and makes exactly the same flow payment as when he
sends $a_x$ under the original mechanism. Hence truthful play under
$\Gamma^D$ generates the same meeting process, matching behavior,
steady-state unmatched fractions, and platform revenue. Market clearing
is therefore preserved.

It remains to establish sequential incentive compatibility. Consider any
admissible deviation by type $x$ under the direct mechanism. At every
unmatched history, such a deviation selects a report in $X$ or exits,
possibly as a function of the agent's entire private history and possibly
using randomization. Construct a deviation under the original mechanism
by replacing every report $\hat x$ with message $a_{\hat x}$ and leaving
exit unchanged.

At every history the two deviations induce exactly the same search
distribution and flow payment. They consequently generate the same
distribution over subsequent meetings and matches and give the agent the
same expected payoff. If the deviation were profitable under
$\Gamma^D$, its counterpart would therefore be profitable under
$\Gamma$. This contradicts the sequential optimality of sending $a_x$
whenever unmatched in the original stationary implementation.

\end{proof}

Combining Propositions \ref{P:stationary-reduction} and
\ref{P:stationary-RP} gives the mechanism class used in the main text. That is, it is without loss for the designer to restrict attention to platforms $P$ and associated DSE satisfying \eqref{E:dynamicIR} and \eqref{E:dynamicIC}.

\begin{rem}
The qualification ``within the class of stationary mechanisms'' is
important. Proposition \ref{P:stationary-reduction} eliminates auxiliary randomization in the meeting rule. It does not claim that a lottery which creates an additional persistent contract state can be purified, nor does Proposition \ref{P:stationary-RP} claim that stationary mechanisms are without loss relative to arbitrary history-dependent dynamic contracts. Those richer contracts are outside the baseline mechanism class. Within the baseline class, however, a search distribution already incorporates all relevant randomization over potential partner types.
\end{rem}
\section{Numerical Examples Without the No-Holdout Condition}
\label{A:holdoutexamples}

This appendix provides two numerical examples illustrating what can occur when the no-holdout condition is violated. The first shows that the revenue-maximizing platform may introduce mismatch even when assortative search is efficient. The second shows that positive mismatch may itself be efficient.

\subsection{Profitable but Inefficient Mismatch}

Let $\pi=1/2$ and
\begin{align*}
\rho=1,\qquad
\alpha=10^{-6},\qquad
r=0.004999,
\end{align*}
so that $\theta=100$. Let
\begin{align*}
f_{\lo\lo}=0.01,\qquad
f_{\lo\hi}=0.5,\qquad
f_{\hi\hi}=1.
\end{align*}
These primitives satisfy strict monotonicity and strict supermodularity, but violate the no-holdout condition.

Under assortative search,
\begin{align*}
u^0\simeq0.0009995,\
w_\lo^0\simeq0.000833,\
w_\hi^0\simeq0.083299.
\end{align*}
Assortative full inclusion generates revenue approximately $0.01980$, while exclusion of the low type generates revenue approximately $0.04165$.

Now consider $\epsilon=0.45$. There is a DSE in which low-low meetings are rejected, while low-high and high-high meetings are accepted. The unmatched fractions and reservation wages are
\begin{align*}
u_\lo\simeq0.005037,
u_\hi\simeq0.000219,\
w_\lo\simeq0.006685,
w_\hi\simeq0.154917.
\end{align*}
The persistent-deviation values are
\begin{align*}
\widetilde w_{\hi,\lo}\simeq0.038819,
\qquad
\widetilde w_{\lo,\hi}\simeq0.001553.
\end{align*}
The corresponding allocation is implementable. Revenue-maximizing transfers are
\begin{align*}
t_\lo\simeq0.006685,
\qquad
t_\hi\simeq0.122782,
\end{align*}
which generate revenue approximately $0.06473$. Thus, this mismatched platform strictly dominates both assortative alternatives, and every revenue-maximizing platform must introduce mismatch.

The same primitives yield the opposite welfare comparison. Assortative search generates steady-state output
\begin{align*}
\mathcal W(0)\simeq0.252248.
\end{align*}
A direct numerical maximization over the low-low-rejection branch gives maximal output approximately $0.249953$. Proposition \ref{P:efficiency} implies that positive mismatch cannot improve output in the remaining equilibrium regimes. Hence, assortative search is efficient for these primitives even though every revenue-maximizing platform introduces mismatch. The optimal platform therefore need not be efficient.

\subsection{Efficient Mismatch}

Positive mismatch can also be socially desirable. Let $\pi=1/2$, $\rho=1$, $\alpha=0.27$, and $r=0.01$, and let
\begin{align*}
f_{\lo\lo}=0.08,\qquad
f_{\lo\hi}=1,\qquad
f_{\hi\hi}=1.95.
\end{align*}
Again, the production function is strictly monotone and strictly supermodular, while the No-Holdout Condition is violated.

Under assortative search, steady-state output is approximately
\begin{align*}
\mathcal W(0)\simeq0.303553.
\end{align*}
Output is maximized at approximately
\begin{align*}
\epsilon^{eff}\simeq0.2727.
\end{align*}
At this mismatch rate,
\begin{align*}
u_\lo\simeq0.57879,
\qquad
u_\hi\simeq0.36026,
\end{align*}
and steady-state output is approximately
\begin{align*}
\mathcal W(\epsilon^{eff})\simeq0.317136.
\end{align*}
Hence, the assortative platform is not efficient.

\section{Proofs}\label{A:proofs}

Throughout this appendix, for any full-inclusion platform write
\begin{align}
    a=\frac{\epsilon}{1-\pi},
    \qquad
    b=\frac{\epsilon}{\pi}.
    \label{E:Aab}
\end{align}
Thus $a=g_{\lo}(\hi)$ and $b=g_{\hi}(\lo)$, and market clearing implies
\begin{align}
    (1-\pi)a=\pi b=\epsilon.
    \label{E:Amc}
\end{align}

\begin{proofof}{\bf Proposition \ref{P:assortDSE}}
On an assortative platform, type $x$ meets only type $x$. An own-type meeting cannot be rejected in equilibrium: if it were, the Bellman equation would imply $w_x=0$, but then $f_{xx}-2w_x=f_{xx}>0$, so the meeting would be strictly acceptable. Hence $x\in\mathcal M(x)$.

The steady-state balance equation is therefore
\begin{align*}
    \alpha(1-u_x)=\rho u_x^2,
\end{align*}
whose unique solution in $(0,1)$ is $u_x=u^0$. The Bellman equation then becomes
\begin{align*}
    w_x=\theta u^0(f_{xx}-2w_x),
\end{align*}
so
\begin{align*}
    w_x=\frac{\theta u^0}{1+2\theta u^0}f_{xx}=w_x^0.
\end{align*}
Thus the assortative DSE is unique.\hfill$\square$
\end{proofof}

\begin{proofof}{\bf Lemma \ref{L:regimes}}
First, neither type can reject every possible match. If $\mathcal M(x)=\emptyset$, then the Bellman equation implies $w_x=0$. But then
\begin{align*}
    f_{xx}-2w_x=f_{xx}>0,
\end{align*}
so an own-type match is strictly acceptable, a contradiction.

Suppose first that cross-type meetings are rejected. Since neither type can reject every possible match, each type must accept its own type. That is, $\mathcal{M}(x)=\{x\}$, which is the selective regime.

Now suppose that cross-type meetings are accepted. The two types cannot both reject own-type meetings. Otherwise,
\begin{align*}
    f_{\lo\lo}\leq 2w_{\lo},
    \qquad
    f_{\hi\hi}\leq 2w_{\hi},
\end{align*}
while cross-type acceptance implies
\begin{align*}
    f_{\lo\hi}>w_{\lo}+w_{\hi}.
\end{align*}
It would follow that
\begin{align*}
    f_{\lo\lo}+f_{\hi\hi}
    \leq 2(w_{\lo}+w_{\hi})
    <2f_{\lo\hi},
\end{align*}
contradicting strict supermodularity.

It remains only to rule out the case in which $\mathcal{M}(\lo)=\{\lo,\hi\}$ and $\mathcal{M}(\hi)=\{\lo\}$. Suppose, toward a contradiction, that this was the case. The steady-state conditions are then
\begin{align*}
    \alpha(1-u_{\lo})
    &=
    \rho u_{\lo}
    \left[
        g_{\lo}(\lo)u_{\lo}
        +g_{\lo}(\hi)u_{\hi}
    \right],\\
    \alpha(1-u_{\hi})
    &=
    \rho u_{\hi}g_{\hi}(\lo)u_{\lo}.
\end{align*}
Subtracting the first equation from the second gives
\begin{align*}
    \alpha(u_{\lo}-u_{\hi})
    =
    \rho u_{\lo}
    \left[
        \bigl(g_{\hi}(\lo)-g_{\lo}(\hi)\bigr)u_{\hi}
        -g_{\lo}(\lo)u_{\lo}
    \right].
\end{align*}
I claim that $u_{\lo}\leq u_{\hi}$. Suppose instead that $u_{\lo}>u_{\hi}$. If $g_{\hi}(\lo)\leq g_{\lo}(\hi)$, the right-hand side is nonpositive, contradicting the strictly positive left-hand side. If $g_{\hi}(\lo)>g_{\lo}(\hi)$, then
\begin{align*}
    &\bigl(g_{\hi}(\lo)-g_{\lo}(\hi)\bigr)u_{\hi}
        -g_{\lo}(\lo)u_{\lo}\\
    &\qquad<
    \bigl[
        g_{\hi}(\lo)-g_{\lo}(\hi)-g_{\lo}(\lo)
    \bigr]u_{\lo}\\
    &\qquad=
    \bigl[g_{\hi}(\lo)-1\bigr]u_{\lo}
    \leq0,
\end{align*}
again a contradiction. Hence $u_{\lo}\leq u_{\hi}$.

Dividing the two steady-state conditions by $\rho u_{\lo}$ and $\rho u_{\hi}$, respectively, and using that $\frac{\alpha(1-u)}{\rho u}$ is decreasing in $u$, I obtain
\begin{align*}
    g_{\lo}(\lo)u_{\lo}
    +g_{\lo}(\hi)u_{\hi}
    \geq
    g_{\hi}(\lo)u_{\lo}.
\end{align*}
Since high-high meetings are rejected,
\begin{align*}
    f_{\hi\hi}-2w_{\hi}\leq0.
\end{align*}
Strict supermodularity therefore implies
\begin{align*}
    f_{\lo\lo}-2w_{\lo}
    >
    2\bigl(f_{\lo\hi}-w_{\lo}-w_{\hi}\bigr)
    >
    f_{\lo\hi}-w_{\lo}-w_{\hi},
\end{align*}
where the second inequality follows from cross-type acceptance. The Bellman equations consequently imply
\begin{align*}
    w_{\lo}
    &=
    \theta g_{\lo}(\lo)u_{\lo}
        (f_{\lo\lo}-2w_{\lo})
    +
    \theta g_{\lo}(\hi)u_{\hi}
        (f_{\lo\hi}-w_{\lo}-w_{\hi})\\
    &\geq
    \theta
    \left[
        g_{\lo}(\lo)u_{\lo}
        +g_{\lo}(\hi)u_{\hi}
    \right]
    (f_{\lo\hi}-w_{\lo}-w_{\hi})\\
    &\geq
    \theta g_{\hi}(\lo)u_{\lo}
    (f_{\lo\hi}-w_{\lo}-w_{\hi})
    =
    w_{\hi}.
\end{align*}
Cross-type acceptance then gives
\begin{align*}
    f_{\lo\hi}>w_{\lo}+w_{\hi}\geq2w_{\hi}.
\end{align*}
Since $f_{\hi\hi}>f_{\lo\hi}$, this implies
\begin{align*}
    f_{\hi\hi}-2w_{\hi}>0,
\end{align*}
contradicting $\hi\not\in\mathcal{M}(\hi)$.

Therefore, when cross-type meetings are accepted, either both own-type meetings are accepted, giving the accepting regime, or only low-low meetings are rejected, giving the holdout regime. These three cases exhaust all possibilities.\hfill$\square$
\end{proofof}

\begin{proofof}{\bf Lemma \ref{L:mon}}
Fix $\epsilon',\epsilon\in[0,\bar\epsilon]$ with $\epsilon'>\epsilon$. Consider first the low type. The balance equations at $\epsilon$ and $\epsilon'$ are
\begin{align*}
    \alpha(1-u_\lo^S(\epsilon))
    &=
    \rho\left(1-\frac{\epsilon}{1-\pi}\right)
    (u_\lo^S(\epsilon))^2,\\
    \alpha(1-u_\lo^S(\epsilon'))
    &=
    \rho\left(1-\frac{\epsilon'}{1-\pi}\right)
    (u_\lo^S(\epsilon'))^2.
\end{align*}
Suppose, toward a contradiction, that
\begin{align*}
    u_\lo^S(\epsilon')\leq u_\lo^S(\epsilon).
\end{align*}
Then
\begin{align*}
    \alpha(1-u_\lo^S(\epsilon'))
    \geq
    \alpha(1-u_\lo^S(\epsilon)),
\end{align*}
while
\begin{align*}
    \rho\left(1-\frac{\epsilon'}{1-\pi}\right)
    (u_\lo^S(\epsilon'))^2
    &<
    \rho\left(1-\frac{\epsilon}{1-\pi}\right)
    (u_\lo^S(\epsilon))^2,
\end{align*}
contradicting the two balance equations. Hence $u_\lo^S(\epsilon')>u_\lo^S(\epsilon)$. The same argument gives $u_\hi^S(\epsilon')>u_\hi^S(\epsilon)$.

For reservation wages, the low-type balance equation can be rearranged as
\begin{align*}
    \left(1-\frac{\epsilon}{1-\pi}\right)u_\lo^S(\epsilon)
    =
    \frac{\alpha}{\rho}
    \frac{1-u_\lo^S(\epsilon)}{u_\lo^S(\epsilon)}.
\end{align*}
Since $u_\lo^S(\epsilon')>u_\lo^S(\epsilon)$,
\begin{align*}
    \left(1-\frac{\epsilon'}{1-\pi}\right)u_\lo^S(\epsilon')
    <
    \left(1-\frac{\epsilon}{1-\pi}\right)u_\lo^S(\epsilon).
\end{align*}
The Bellman equation implies
\begin{align*}
    w_\lo^S(\epsilon)
    =
    \frac{
        \theta\left(1-\frac{\epsilon}{1-\pi}\right)
        u_\lo^S(\epsilon)
    }{
        1+
        2\theta\left(1-\frac{\epsilon}{1-\pi}\right)
        u_\lo^S(\epsilon)
    }
    f_{\lo\lo}.
\end{align*}
The right-hand side is strictly increasing in
$\left(1-\frac{\epsilon}{1-\pi}\right)u_\lo^S(\epsilon)$.
Therefore $w_\lo^S(\epsilon')<w_\lo^S(\epsilon)$. Applying the same argument to the high type gives $w_\hi^S(\epsilon')<w_\hi^S(\epsilon)$.\hfill$\square$
\end{proofof}

\begin{proofof}{\bf Lemma \ref{L:uniqueacc}}
Suppose $(w,\mathcal M,u)$ is an accepting DSE. Then every meeting is
accepted, so the balance equations are
\begin{align*}
    \alpha(1-u_{\lo})
    &=
    \rho u_{\lo}
    \left[
        \left(1-\frac{\epsilon}{1-\pi}\right)u_{\lo}
        +\frac{\epsilon}{1-\pi}u_{\hi}
    \right],\\
    \alpha(1-u_{\hi})
    &=
    \rho u_{\hi}
    \left[
        \frac{\epsilon}{\pi}u_{\lo}
        +\left(1-\frac{\epsilon}{\pi}\right)u_{\hi}
    \right].
\end{align*}
Dividing each equation by $\rho u_x$ and subtracting, if $u_{\lo}\neq u_{\hi}$ I obtain
\begin{align*}
    \frac{\alpha}{\rho u_{\lo}u_{\hi}}
    =
    \frac{\epsilon}{1-\pi}
    +\frac{\epsilon}{\pi}
    -1.
\end{align*}
In particular, the right-hand side must be strictly positive. Substituting
this equality into the low-type balance equation gives
\begin{align*}
    \left[
        \frac{\epsilon}{\pi}-1
        -
        \left(
            \frac{\epsilon}{1-\pi}
            +\frac{\epsilon}{\pi}
            -1
        \right)u_{\lo}
    \right]u_{\hi}
    =
    \left(1-\frac{\epsilon}{1-\pi}\right)u_{\lo}.
\end{align*}
The left-hand side is strictly negative, whereas the right-hand side is nonnegative, a contradiction. Hence $u_\lo=u_\hi$. Either balance equation then reduces to
\begin{align*}
    \alpha(1-u_x)=\rho u_x^2,
\end{align*}
whose unique (non-negative) solution is $u_x=u^0$. Therefore $u_x=u^0$ for $x\in\{\lo,\hi\}$.

It remains to show that the reservation wages are uniquely determined. Since every meeting is accepted, the Bellman equations are
\begin{align*}
    w_{\lo}
    &=
    \theta u^0
    \left[
        \left(1-\frac{\epsilon}{1-\pi}\right)
        (f_{\lo\lo}-2w_{\lo})
        +
        \frac{\epsilon}{1-\pi}
        (f_{\lo\hi}-w_{\lo}-w_{\hi})
    \right],\\
    w_{\hi}
    &=
    \theta u^0
    \left[
        \frac{\epsilon}{\pi}
        (f_{\lo\hi}-w_{\lo}-w_{\hi})
        +
        \left(1-\frac{\epsilon}{\pi}\right)
        (f_{\hi\hi}-2w_{\hi})
    \right].
\end{align*}
This is a system of two linear equations in $w_{\lo}$ and $w_{\hi}$.
Its determinant is
\begin{align*}
    (1+2\theta u^0)
    \left[
        1+2\theta u^0
        -
        \theta u^0
        \left(
            \frac{\epsilon}{1-\pi}
            +\frac{\epsilon}{\pi}
        \right)
    \right].
\end{align*}
Since
\begin{align*}
    \frac{\epsilon}{1-\pi}\leq1,
    \qquad
    \frac{\epsilon}{\pi}\leq1,
\end{align*}
the second factor is at least $1$. Thus the determinant is strictly positive, and the Bellman equations have a unique solution. Hence, if an accepting DSE exists for $\epsilon$, it is unique.\hfill$\square$
\end{proofof}

\begin{proofof}{\bf Proposition \ref{P:regions}}
Since $f_{\lo\hi}-w_\lo^S(\epsilon)-w_\hi^S(\epsilon)$ is strictly increasing in $\epsilon$ by Lemma \ref{L:mon}, there exists a threshold $\epsilon^D$ such that a selective equilibrium exists if and only if $\epsilon\leq\epsilon^D$.

I next establish the analogous threshold result for the accepting regime. By Lemma \ref{L:uniqueacc}, the accepting candidate is unique for every
$\epsilon$. Moreover,
\begin{align*}
    u_\lo^A(\epsilon)=u_\hi^A(\epsilon)=u^0,
\end{align*}
and its reservation wages uniquely solve
\begin{align*}
    w_\lo^A(\epsilon)
    &=
    \theta u^0
    \left[
        \left(1-\frac{\epsilon}{1-\pi}\right)
        (f_{\lo\lo}-2w_\lo^A(\epsilon))
        +
        \frac{\epsilon}{1-\pi}
        (f_{\lo\hi}-w_\lo^A(\epsilon)-w_\hi^A(\epsilon))
    \right],\\
    w_\hi^A(\epsilon)
    &=
    \theta u^0
    \left[
        \frac{\epsilon}{\pi}
        (f_{\lo\hi}-w_\lo^A(\epsilon)-w_\hi^A(\epsilon))
        +
        \left(1-\frac{\epsilon}{\pi}\right)
        (f_{\hi\hi}-2w_\hi^A(\epsilon))
    \right].
\end{align*}

I first show that, under the no-holdout condition, this candidate is an accepting equilibrium if and only if
\begin{align}\label{E:P21}
    f_{\lo\hi}-w_\lo^A(\epsilon)-w_\hi^A(\epsilon)>0.
\end{align}
Necessity is immediate. For sufficiency, observe that if $\lo\not\in\mathcal{M}(\lo)$, then $f_{\lo\lo}-2w_\lo^A(\epsilon)\leq0$. Strict supermodularity and \eqref{E:P21} would then imply $f_{\hi\hi}-2w_\hi^A(\epsilon)>0$. Therefore $w_\hi^A(\epsilon)>0$. The low-type Bellman equation
would therefore give
\begin{align*}
    w_\lo^A(\epsilon)
    &\leq
    \theta u^0\frac{\epsilon}{1-\pi}
    \left(
        f_{\lo\hi}
        -w_\lo^A(\epsilon)-w_\hi^A(\epsilon)
    \right)\\
    &<
    \theta u^0
    \left(
        f_{\lo\hi}-w_\lo^A(\epsilon)
    \right).
\end{align*}
Hence
\begin{align*}
    w_\lo^A(\epsilon)
    <
    \frac{\theta u^0}{1+\theta u^0}f_{\lo\hi}.
\end{align*}
Together with $f_{\lo\lo}\leq2w_\lo^A(\epsilon)$, this contradicts the no-holdout condition. Thus $\lo\in\mathcal{M}(\lo)$.

Suppose instead that high-high meetings were not acceptable. Strict supermodularity and \eqref{E:P21} then imply
\begin{align*}
    f_{\lo\lo}-2w_\lo^A(\epsilon)
    >
    2\left(
        f_{\lo\hi}
        -w_\lo^A(\epsilon)-w_\hi^A(\epsilon)
    \right).
\end{align*}
The Bellman equations consequently give
\begin{align*}
    w_\lo^A(\epsilon)
    &\geq
    \theta u^0
    \left(2-\frac{\epsilon}{1-\pi}\right)
    \left(
        f_{\lo\hi}
        -w_\lo^A(\epsilon)-w_\hi^A(\epsilon)
    \right),\\
    w_\hi^A(\epsilon)
    &\leq
    \theta u^0\frac{\epsilon}{\pi}
    \left(
        f_{\lo\hi}
        -w_\lo^A(\epsilon)-w_\hi^A(\epsilon)
    \right).
\end{align*}
Since $2-\frac{\epsilon}{1-\pi}\geq1\geq\frac{\epsilon}{\pi}$, we have $w_\lo^A(\epsilon)\geq w_\hi^A(\epsilon)$. Equation \eqref{E:P21} then implies $f_{\lo\hi}>2w_\hi^A(\epsilon)$.Since $f_{\hi\hi}>f_{\lo\hi}$, then $h\in\mathcal{M}(h)$, a contradiction. Since $\lo\in\mathcal{M}(\lo)$ and $\hi\in\mathcal{M}(\hi)$, by equation \eqref{E:P21} this is an accepting DSE.

It therefore remains to characterize the sign of \eqref{E:P21}. The linear system above has which characterizes the equilibrium a strictly positive determinant by the proof of Lemma \ref{L:uniqueacc}, so $w_\lo^A(\epsilon)$ and $w_\hi^A(\epsilon)$ are
continuous and differentiable in $\epsilon$. Consider any $\epsilon$ such that the left hand side of \eqref{E:P21} equals zero. At such a point, the Bellman equations imply
\begin{align*}
    f_{\lo\lo}-2w_\lo^A(\epsilon)>0,
    \qquad
    f_{\hi\hi}-2w_\hi^A(\epsilon)>0.
\end{align*}
Differentiating the two Bellman equations at $\epsilon$ gives
\begin{align*}
&\left[
    1+
    2\theta u^0
    \left(1-\frac{\epsilon}{1-\pi}\right)
\right]
\frac{d w_\lo^A(\epsilon)}{d\epsilon}\\
&\qquad=
-\frac{\theta u^0}{1-\pi}
    (f_{\lo\lo}-2w_\lo^A(\epsilon))
+
\theta u^0\frac{\epsilon}{1-\pi}
\frac{d}{d\epsilon}
\left(
    f_{\lo\hi}
    -w_\lo^A(\epsilon)-w_\hi^A(\epsilon)
\right),
\end{align*}
and
\begin{align*}
&\left[
    1+
    2\theta u^0
    \left(1-\frac{\epsilon}{\pi}\right)
\right]
\frac{d w_\hi^A(\epsilon)}{d\epsilon}\\
&\qquad=
-\frac{\theta u^0}{\pi}
    (f_{\hi\hi}-2w_\hi^A(\epsilon))
+
\theta u^0\frac{\epsilon}{\pi}
\frac{d}{d\epsilon}
\left(
    f_{\lo\hi}
    -w_\lo^A(\epsilon)-w_\hi^A(\epsilon)
\right).
\end{align*}
Since
\begin{align*}
    \frac{d}{d\epsilon}
    \left(
        f_{\lo\hi}
        -w_\lo^A(\epsilon)-w_\hi^A(\epsilon)
    \right)
    =
    -\frac{d w_\lo^A(\epsilon)}{d\epsilon}
    -\frac{d w_\hi^A(\epsilon)}{d\epsilon},
\end{align*}
these equations imply
\begin{align*}
&\frac{d}{d\epsilon}
\left(
    f_{\lo\hi}
    -w_\lo^A(\epsilon)-w_\hi^A(\epsilon)
\right)
\Bigg[
1+
\frac{
    \theta u^0\frac{\epsilon}{1-\pi}
}{
    1+
    2\theta u^0
    \left(1-\frac{\epsilon}{1-\pi}\right)
}
+
\frac{
    \theta u^0\frac{\epsilon}{\pi}
}{
    1+
    2\theta u^0
    \left(1-\frac{\epsilon}{\pi}\right)
}
\Bigg]\\
&\qquad=
\frac{
    \frac{\theta u^0}{1-\pi}
    (f_{\lo\lo}-2w_\lo^A(\epsilon))
}{
    1+
    2\theta u^0
    \left(1-\frac{\epsilon}{1-\pi}\right)
}
+
\frac{
    \frac{\theta u^0}{\pi}
    (f_{\hi\hi}-2w_\hi^A(\epsilon))
}{
    1+
    2\theta u^0
    \left(1-\frac{\epsilon}{\pi}\right)
}
>0.
\end{align*}
Thus every zero of $f_{\lo\hi}-w_\lo^A(\epsilon)-w_h^A(\epsilon)$ is crossed from below. Consequently it has at most one zero, and there exists a threshold $\epsilon^A$ such that an accepting equilibrium
exists if and only if $\epsilon>\epsilon^A$.

It remains to compare $\epsilon^A$ and $\epsilon^D$. I first record a useful comparison. Suppose
\begin{align}\label{E:P23}
    f_{\lo\hi}-w_\lo^A(\epsilon)-w_\hi^A(\epsilon)\leq0.
\end{align}
The low-type accepting Bellman equation then implies
\begin{align*}
    w_\lo^A(\epsilon)
    \leq
    \theta u^0
    \left(1-\frac{\epsilon}{1-\pi}\right)
    (f_{\lo\lo}-2w_\lo^A(\epsilon)).
\end{align*}
Hence
\begin{align*}
    w_\lo^A(\epsilon)
    \leq
    \frac{
        \theta u^0
        \left(1-\frac{\epsilon}{1-\pi}\right)
    }{
        1+
        2\theta u^0
        \left(1-\frac{\epsilon}{1-\pi}\right)
    }
    f_{\lo\lo}.
\end{align*}
By contrast,
\begin{align*}
    w_\lo^S(\epsilon)
    =
    \frac{
        \theta u_\lo^S(\epsilon)
        \left(1-\frac{\epsilon}{1-\pi}\right)
    }{
        1+
        2\theta u_\lo^S(\epsilon)
        \left(1-\frac{\epsilon}{1-\pi}\right)
    }
    f_{\lo\lo}.
\end{align*}
For $\epsilon>0$, the selective balance equation implies
$u_\lo^S(\epsilon)>u^0$. Therefore $w_\lo^S(\epsilon)> w_\lo^A(\epsilon)$, with strict inequality whenever $1-\epsilon/(1-\pi)>0$. The same argument gives $w_\hi^S(\epsilon)>w_\hi^A(\epsilon)$ with strict inequality whenever $1-\epsilon/\pi>0$. Thus, for every $\epsilon>0$ satisfying \eqref{E:P23}
\begin{align*}
    f_{\lo\hi}
    -w_\lo^S(\epsilon)-w_\hi^S(\epsilon)
    <
    f_{\lo\hi}
    -w_\lo^A(\epsilon)-w_\hi^A(\epsilon)
    \leq0.
\end{align*}
Hence whenever the accepting equilibrium has not yet become viable, the selective equilibrium still exists.

In particular, if $\epsilon^A>0$, then at $\epsilon=\epsilon^A$,
\begin{align*}
    f_{\lo\hi}
    -w_\lo^A(\epsilon^A)-w_\hi^A(\epsilon^A)=0
\end{align*}
while
\begin{align*}
    f_{\lo\hi}
    -w_\lo^S(\epsilon^A)-w_\hi^S(\epsilon^A)<0.
\end{align*}
Since the latter expression is strictly increasing in $\epsilon$, its zero must occur strictly after $\epsilon^A$. Therefore $\epsilon^A<\epsilon^D$.

If the accepting or selective regime exists for every feasible $\epsilon$, or for no feasible $\epsilon$, the corresponding threshold can be placed outside $[0,\bar\epsilon]$ and the same ordering can be
maintained. The only case in which equality can occur is when
\begin{align*}
    f_{\lo\hi}-w_\lo^0-w_\hi^0=0,
\end{align*}
for which $\epsilon^A=\epsilon^D=0$.\hfill$\square$
\end{proofof}

\begin{lem}[No-Holdout Condition]\label{L:noholdout}
If the no-holdout condition is satisfied, then low-low meetings are accepted in every full-inclusion DSE.
\end{lem}

\begin{proof}
Suppose to the contrary that there is a full-inclusion DSE with $\lo\not\in\mathcal{M}(\lo)$. Lemma \ref{L:regimes} implies that $\mathcal{M}(\lo)=\{\hi\}$ and $\mathcal{M}(\hi)=\{\lo,\hi\}$. The balance equations are
\begin{align*}
    \alpha(1-u_{\lo})&=\rho u_{\lo}au_{\hi},\\
    \alpha(1-u_{\hi})&=\rho u_{\hi}
    \left[bu_{\lo}+(1-b)u_{\hi}\right].
\end{align*}
I first show that $u_{\lo}\geq u_{\hi}$. Subtracting the first equation from the second gives
\begin{align}
    \alpha(u_{\lo}-u_{\hi})
    =
    \rho u_{\hi}\left[(b-a)u_{\lo}+(1-b)u_{\hi}\right].
    \label{E:Arejdiff}
\end{align}
Suppose $u_{\lo}<u_{\hi}$. If $b\geq a$, the right-hand side is nonnegative. If $b<a$, then
\begin{align*}
    (b-a)u_{\lo}+(1-b)u_{\hi}
    >
    (b-a)u_{\hi}+(1-b)u_{\hi}
    =(1-a)u_{\hi}\geq0.
\end{align*}
Either case contradicts \eqref{E:Arejdiff}. Hence $u_{\lo}\geq u_{\hi}$.

The high-type balance equation now implies
\begin{align*}
    \frac{\alpha(1-u_{\hi})}{\rho u_{\hi}}
    =bu_{\lo}+(1-b)u_{\hi}
    \geq u_{\hi}.
\end{align*}
Since $u^0$ uniquely solves
\begin{align*}
    \frac{\alpha(1-u^0)}{\rho u^0}=u^0,
\end{align*}
and the left-hand side is decreasing in its argument, it follows that $u_{\hi}\leq u^0$. Since $a\leq1$, $au_{\hi}\leq u^0$.

The low type matches only with high types, so
\begin{align*}
    w_{\lo}
    &=\theta au_{\hi}
    (f_{\lo\hi}-w_{\lo}-w_{\hi})\\
    &<\theta au_{\hi}(f_{\lo\hi}-w_{\lo}).
\end{align*}
Therefore
\begin{align*}
    w_{\lo}
    <\frac{\theta u^0}{1+\theta u^0}f_{\lo\hi}.
\end{align*}
Rejection of low-low meetings requires $f_{\lo\lo}\leq2w_{\lo}$, contradicting the no-holdout condition.
\end{proof}

\begin{lem}[Gross-Value Dominance]\label{L:gross}
If a full-inclusion DSE satisfies $\lo\in\mathcal{M}(\lo)$, then
\begin{align}
    (1-\pi)w_{\lo}+\pi w_{\hi}
    \leq
    (1-\pi)w_{\lo}^0+\pi w_{\hi}^0,
    \label{E:Agross}
\end{align}
with strict inequality whenever $\epsilon>0$.
\end{lem}

\begin{proof}
At $\epsilon=0$, Proposition \ref{P:assortDSE} gives equality. Now suppose $\epsilon>0$. By Lemma \ref{L:regimes}, the DSE is either accepting or selective.

Suppose the DSE is accepting. The balance equations, divided by $\rho u_x$, are
\begin{align*}
    \frac{\alpha(1-u_{\lo})}{\rho u_{\lo}}
    &=(1-a)u_{\lo}+au_{\hi},\\
    \frac{\alpha(1-u_{\hi})}{\rho u_{\hi}}
    &=bu_{\lo}+(1-b)u_{\hi}.
\end{align*}
If $u_{\lo}\neq u_{\hi}$, subtracting the equations gives
\begin{align}
    \frac{\alpha}{\rho u_{\lo}u_{\hi}}
    =a+b-1>0.
    \label{E:Afulldiff}
\end{align}
Substituting \eqref{E:Afulldiff} into the low-type balance equation gives
\begin{align*}
    \left[b-1-(a+b-1)u_{\lo}\right]u_{\hi}
    =(1-a)u_{\lo}.
\end{align*}
The left-hand side is strictly negative whereas the right-hand side is nonnegative, a contradiction. Hence $u_{\lo}=u_{\hi}$, and either balance equation gives $u_{\lo}=u_{\hi}=u^0$. Multiply the low Bellman equation by $1-\pi$, multiply the high Bellman equation by $\pi$, and add. Using \eqref{E:Amc} gives
\begin{align*}
    &(1-\pi)w_{\lo}+\pi w_{\hi}\nonumber
    &\qquad=
    \frac{\theta u^0}{1+2\theta u^0}
    \left[
    (1-\pi)f_{\lo\lo}+\pi f_{\hi\hi}
    -\epsilon(f_{\lo\lo}+f_{\hi\hi}-2f_{\lo\hi})
    \right].
\end{align*}
Strict supermodularity makes this strictly smaller than its value at $\epsilon=0$.

Now suppose the equilibrium is selective. The balance equations reduce to
\begin{align}
    \alpha(1-u_{\lo})&=\rho(1-a)u_{\lo}^2,\nonumber\\
    \alpha(1-u_{\hi})&=\rho(1-b)u_{\hi}^2.
    \label{E:Adiagbal}
\end{align}
If $a<1$, then $u_{\lo}>u^0$ and
\begin{align*}
    (1-a)u_{\lo}
    =\frac{\alpha(1-u_{\lo})}{\rho u_{\lo}}
    <\frac{\alpha(1-u^0)}{\rho u^0}=u^0;
\end{align*}
if $a=1$, the left-hand side is zero. The same argument gives $(1-b)u_{\hi}<u^0$. The Bellman equations are
\begin{align*}
    w_{\lo}&=\theta(1-a)u_{\lo}(f_{\lo\lo}-2w_{\lo}),\\
    w_{\hi}&=\theta(1-b)u_{\hi}(f_{\hi\hi}-2w_{\hi}),
\end{align*}
so $w_{\lo}<w_{\lo}^0$ and $w_{\hi}<w_{\hi}^0$. This proves \eqref{E:Agross}.\hfill$\square$
\end{proof}

\begin{proofof}{\bf Proposition \ref{P:efficiency}}
Suppose $\epsilon>0$ and fix a DSE in which low-low meetings are accepted. By Lemma \ref{L:regimes}, the DSE is either accepting or selective.

Suppose the DSE is accepting. The proof of Lemma \ref{L:gross} shows that $u_{\lo}=u_{\hi}=u^0$. The steady-state masses of low-low, low-high, and high-high matches are, respectively,
\begin{align*}
    \frac{\rho}{2\alpha}(1-\pi-\epsilon)(u^0)^2,
    \qquad
    \frac{\rho}{\alpha}\epsilon(u^0)^2,
    \qquad
    \frac{\rho}{2\alpha}(\pi-\epsilon)(u^0)^2.
\end{align*}
Thus
\begin{align}
    \mathcal W(\epsilon)
    =\frac{\rho(u^0)^2}{2\alpha}
    \left[
    (1-\pi)f_{\lo\lo}+\pi f_{\hi\hi}
    -\epsilon(f_{\lo\lo}+f_{\hi\hi}-2f_{\lo\hi})
    \right].
    \label{E:Awelfareall}
\end{align}
which is strictly decreasing in $\epsilon$ by supermodularity.

Suppose instead that the DSE is selective. The balance equations are \eqref{E:Adiagbal}. For $\epsilon>0$, the proof of Lemma \ref{L:gross} shows that $u_{\lo}>u^0$ and $u_{\hi}>u^0$. Since every realized match is an own-type match, steady-state output is
\begin{align*}
    \mathcal W(\epsilon)
    =\frac12\left[
    (1-\pi)(1-u_{\lo})f_{\lo\lo}
    +\pi(1-u_{\hi})f_{\hi\hi}
    \right]
    <\mathcal W(0).
\end{align*}
This proves the proposition.
\end{proofof}

\begin{proofof}{\bf Corollary \ref{C:efficientsearch}}
This is immediate by Lemma \ref{L:noholdout} and Proposition \ref{P:efficiency}.
\end{proofof}

\begin{proofof}{\bf Proposition \ref{P:dynamicIC}}
Fix a true type $x$ and a report $\hat x$, and define the net value of using $\hat x$ persistently by $\tilde v_{x,\hat x}=\tilde w_{x,\hat x}-t_{\hat x}$. Substituting
$\tilde w_{x,\hat x}=\tilde v_{x,\hat x}+t_{\hat x}$ into the definition of $\tilde{w}_{x,\hat{x}}$ gives
\begin{align*}
    \tilde v_{x,\hat x}
    &=
    -t_{\hat x}
    +
    \theta
    \sum_{y\in\{\lo,\hi\}}
    u_y g_{\hat x}(y)
    \left[
        f_{xy}
        -\tilde v_{x,\hat x}
        -t_{\hat x}
        -w_y
    \right]_+\\
    &=
    \Phi_{x,\hat x}(\tilde v_{x,\hat x}).
\end{align*}
Hence $\tilde v_{x,\hat x}$ is a fixed point of
$\Phi_{x,\hat x}$.

The map $\Phi_{x,\hat x}$ is weakly decreasing in its argument. It follows that
\begin{align*}
    D_{x,\hat x}(v)
    :=
    \Phi_{x,\hat x}(v)-v
\end{align*}
is strictly decreasing. Since $D_{x,\hat x}(\tilde v_{x,\hat x})=0$, we have
\begin{align}
    \Phi_{x,\hat x}(v)\leq v
    \quad\Longleftrightarrow\quad
    v\geq\tilde v_{x,\hat x}.
    \label{E:dynamicEquivalence}
\end{align}

Now evaluate this equivalence at the truthful net continuation value $v_x=w_x-t_x$. Dynamic incentive compatibility, \eqref{E:dynamicIC}, requires $\Phi_{x,\hat x}(v_x)\leq v_x$. By \eqref{E:dynamicEquivalence}, this holds if and only if $v_x\geq\tilde v_{x,\hat{x}}$. Substituting the definitions of $v_x$ and $\tilde v_{x,\hat x}$ gives
\begin{align*}
    w_x-t_x
    \geq
    \tilde w_{x,\hat x}-t_{\hat x},
\end{align*}
which is \eqref{E:IC}. Dynamic individual rationality
\eqref{E:dynamicIR} is already exactly \eqref{E:IR}.\hfill$\square$
\end{proofof}

\begin{proofof}{\bf Lemma \ref{L:revenue-screening}}
Let $U_x=w_x-t_x$ denote truthful utility. By Proposition \ref{P:dynamicIC}, the downward and upward incentive constraints can be written as
\begin{align*}
    U_{\hi}-U_{\lo}
    &\geq \tilde w_{\hi,\lo}-w_{\lo},\\
    U_{\hi}-U_{\lo}
    &\leq w_{\hi}-\tilde w_{\lo,\hi}.
\end{align*}
For any fixed menu element, replacing the true type $\lo$ by $\hi$ strictly raises every match output. Evaluating the high type's deviation operator at the low type's truthful fixed point therefore gives a value strictly above $w_{\lo}$, so $\tilde w_{\hi,\lo}>w_{\lo}$. Reversing the comparison on the high menu gives $\tilde w_{\lo,\hi}<w_{\hi}$.

The two incentive constraints are jointly feasible if and only if $\tilde w_{\hi,\lo}-w_{\lo}\leq w_{\hi}-\tilde w_{\lo,\hi}$, which is exactly \eqref{E:SCC}. If \eqref{E:SCC} holds, setting
\begin{align*}
    U_{\lo}=0,
    \qquad
    U_{\hi}=\tilde w_{\hi,\lo}-w_{\lo}>0
\end{align*}
satisfies both incentive constraints and individual rationality. These utilities correspond to
\begin{align*}
    t_{\lo}=w_{\lo},
    \qquad
    t_{\hi}=w_{\hi}-(\tilde w_{\hi,\lo}-w_{\lo}),
\end{align*}
proving part (i).

Conversely, feasibility implies that the interval defined by the two incentive constraints is nonempty, so \eqref{E:SCC} must hold. Revenue can be written as
\begin{align*}
    (1-\pi)(w_{\lo}-U_{\lo})+\pi(w_{\hi}-U_{\hi}).
\end{align*}
Because both population shares are positive, revenue is maximized by choosing the smallest feasible utilities. The lower bounds are $U_{\lo}\geq0$ and $U_{\hi}\geq U_{\lo}+\tilde w_{\hi,\lo}-w_{\lo}$, so the revenue-maximizing choice is precisely $U_{\lo}=0$ and $U_{\hi}=\tilde w_{\hi,\lo}-w_{\lo}$. This yields the transfers stated in part (ii).\hfill$\square$
\end{proofof}

\begin{proofof}{\bf Proposition \ref{P:nodist}} The first inequality follows immediately from Lemma \ref{L:gross}. For the rent comparison, write
\begin{align*}
    d=\tilde w_{\hi,\lo}-w_{\lo},
    \qquad
    d^0=\frac{\theta u^0}{1+\theta u^0}
    (f_{\lo\hi}-f_{\lo\lo}).
\end{align*}
By Lemma \ref{L:regimes}, there are two cases: the DSE is accepting or selective.

Suppose first that the DSE is accepting. Lemma \ref{L:gross} gives $u_{\lo}=u_{\hi}=u^0$. I first show that if a deviating high type accepts a low partner, it also accepts a high partner. Let
\begin{align*}
    q=\theta u^0(1-a).
\end{align*}
The truthful low-type Bellman equation contains a nonnegative cross-type term, so
\begin{align*}
    w_{\lo}
    \geq q(f_{\lo\lo}-2w_{\lo})
    \quad\Longrightarrow\quad
    w_{\lo}\geq\frac{q}{1+2q}f_{\lo\lo}.
\end{align*}
If a deviating high type accepted low partners but rejected high partners, its value would satisfy
\begin{align*}
    \tilde w_{\hi,\lo}
    =q(f_{\lo\hi}-\tilde w_{\hi,\lo}-w_{\lo}),
\end{align*}
and hence
\begin{align*}
    \tilde w_{\hi,\lo}
    &=
    \frac{q}{1+q}(f_{\lo\hi}-w_{\lo})\\
    &\leq
    \frac{q}{1+q}
    \left(f_{\lo\hi}-\frac{q}{1+2q}f_{\lo\lo}\right)
    <f_{\lo\hi}-\frac12f_{\lo\lo}.
\end{align*}
The final inequality follows from $f_{\lo\hi}>f_{\lo\lo}$. Strict supermodularity implies
\begin{align*}
    f_{\lo\hi}-\frac12f_{\lo\lo}<\frac12f_{\hi\hi}.
\end{align*}
Since truthful high-high meetings are accepted, $w_{\hi}<f_{\hi\hi}/2$, so
\begin{align*}
    f_{\hi\hi}-w_{\hi}>\frac12f_{\hi\hi}>
    \tilde w_{\hi,\lo}.
\end{align*}
The deviator would therefore accept a high partner, a contradiction.

If the deviating high type rejects a low partner, then
\begin{align*}
    \tilde w_{\hi,\lo}\geq f_{\lo\hi}-w_{\lo},
\end{align*}
and low-low acceptance gives
\begin{align*}
    d\geq f_{\lo\hi}-2w_{\lo}
    \geq f_{\lo\hi}-f_{\lo\lo}>d^0.
\end{align*}
If instead the deviator accepts a low partner, it accepts both partner types. Subtracting the truthful low Bellman equation from the deviation equation gives
\begin{align*}
    d
    =\theta u^0\left[
    (1-a)(f_{\lo\hi}-f_{\lo\lo}-d)
    +a(f_{\hi\hi}-f_{\lo\hi}-d)
    \right].
\end{align*}
Thus
\begin{align}
    d
    =
    \frac{\theta u^0}{1+\theta u^0}
    \left[
    (1-a)(f_{\lo\hi}-f_{\lo\lo})
    +a(f_{\hi\hi}-f_{\lo\hi})
    \right].
    \label{E:Afullrent}
\end{align}
Strict supermodularity implies
\begin{align*}
    f_{\hi\hi}-f_{\lo\hi}>f_{\lo\hi}-f_{\lo\lo},
\end{align*}
so \eqref{E:Afullrent} is at least $d^0$, with strict inequality when $\epsilon>0$.

Now suppose the DSE is selective. Define
\begin{align*}
    q=\theta(1-a)u_{\lo},
    \qquad
    p=\theta au_{\hi}.
\end{align*}
For $\epsilon>0$, \eqref{E:Adiagbal} implies $u_{\lo}>u^0$ and $u_{\hi}>u^0$; if an own-type search probability is zero, the corresponding unmatched fraction is one. Hence
\begin{align}
    q+p
    =\theta\left[(1-a)u_{\lo}+au_{\hi}\right]
    >\theta u^0.
    \label{E:Aqp}
\end{align}

If the deviating high type rejects a low partner, the same argument as above gives $d>f_{\lo\hi}-f_{\lo\lo}>d^0$. If it accepts a low partner, it also accepts a high partner: truthful rejection of a cross-type meeting gives $f_{\lo\hi}-w_{\lo}\leq w_{\hi}$, while truthful acceptance of high-high meetings gives $f_{\hi\hi}-w_{\hi}>w_{\hi}$.

Subtracting the truthful low Bellman equation from the deviation equation therefore gives
\begin{align*}
    d
    =q(f_{\lo\hi}-f_{\lo\lo}-d)
    +p(f_{\hi\hi}-w_{\lo}-w_{\hi}-d).
\end{align*}
Own-type acceptance implies $w_{\lo}<f_{\lo\lo}/2$ and $w_{\hi}<f_{\hi\hi}/2$. Hence strict supermodularity gives
\begin{align*}
    f_{\hi\hi}-w_{\lo}-w_{\hi}
    >
    \frac12(f_{\hi\hi}-f_{\lo\lo})
    >
    f_{\lo\hi}-f_{\lo\lo}.
\end{align*}
It follows that
\begin{align*}
    d>
    \frac{q+p}{1+q+p}(f_{\lo\hi}-f_{\lo\lo})
    >
    d^0,
\end{align*}
where the last inequality follows from \eqref{E:Aqp}. Hence, the rent inequality holds.

By Lemma \ref{L:revenue-screening}, maximal revenue from any implementable full-inclusion allocation is
\begin{align*}
    (1-\pi)w_{\lo}
    +\pi\left[w_{\hi}-(\tilde w_{\hi,\lo}-w_{\lo})\right].
\end{align*}
Therefore every non-assortative full-inclusion platform on which $\lo\in\mathcal{M}(\lo)$ is revenue dominated by the assortative platform.
\end{proofof}

\begin{lem}[Participation and Benchmark Facts]\label{L:participation}
The following statements hold.
\begin{enumerate}[(i)]
    \item An incentive-compatible nonempty platform either includes both types or only the high type.
    \item Assortative full inclusion is implementable, and its high-type information rent is
    \begin{align}
        \tilde w_{\hi,\lo}^0-w_{\lo}^0
        =\frac{\theta u^0}{1+\theta u^0}
        (f_{\lo\hi}-f_{\lo\lo}).
        \label{E:Arent0}
    \end{align}
    \item High-only exclusion is always implementable, with maximal revenue $\pi w_{\hi}^0$.
\end{enumerate}
\end{lem}

\begin{proof}
For part (i), suppose the low type is included while the high type is excluded. Then $g_{\lo}=\delta_{\lo}$ and $w_{\lo}>0$. If a high type chooses the low menu, its deviation value solves
\begin{align*}
    \tilde w_{\hi,\lo}
    =\theta u_{\lo}
    [f_{\lo\hi}-\tilde w_{\hi,\lo}-w_{\lo}]_+.
\end{align*}
Evaluating the deviation operator at $\tilde w_{\hi,\lo}=w_{\lo}$ and using $f_{\lo\hi}>f_{\lo\lo}$ together with
\begin{align*}
    w_{\lo}=\theta u_{\lo}(f_{\lo\lo}-2w_{\lo})
\end{align*}
shows that the operator is strictly larger than $w_{\lo}$. Hence $\tilde w_{\hi,\lo}>w_{\lo}$. Low-type IR requires $t_{\lo}\leq w_{\lo}$, whereas deterring the excluded high type requires $t_{\lo}\geq\tilde w_{\hi,\lo}$, a contradiction.

For part (ii), the high type's value from choosing the low assortative menu satisfies
\begin{align*}
    \tilde w_{\hi,\lo}^0
    =\theta u^0(f_{\lo\hi}-\tilde w_{\hi,\lo}^0-w_{\lo}^0).
\end{align*}
Subtracting
\begin{align*}
    w_{\lo}^0=\theta u^0(f_{\lo\lo}-2w_{\lo}^0)
\end{align*}
gives \eqref{E:Arent0}. If a low type obtains positive value from the high menu, the same calculation gives
\begin{align*}
    \tilde w_{\lo,\hi}^0-w_{\hi}^0
    =\frac{\theta u^0}{1+\theta u^0}
    (f_{\lo\hi}-f_{\hi\hi}).
\end{align*}
Adding the two deviation gaps gives a negative number by strict supermodularity. If the low type obtains zero from the high menu, its deviation gap is $-w_{\hi}^0$, which is smaller still. Thus \eqref{E:SCC} holds, and Lemma \ref{L:revenue-screening} implies that assortative full inclusion is implementable.

For part (iii), if only the high type is included, $g_{\hi}=\delta_{\hi}$ and the high type obtains $w_{\hi}^0$. An excluded low type who chooses the high menu either obtains zero or solves
\begin{align*}
    \tilde w_{\lo,\hi}
    =\theta u^0(f_{\lo\hi}-\tilde w_{\lo,\hi}-w_{\hi}^0).
\end{align*}
In the latter case,
\begin{align*}
    \tilde w_{\lo,\hi}-w_{\hi}^0
    =\frac{\theta u^0}{1+\theta u^0}
    (f_{\lo\hi}-f_{\hi\hi})<0.
\end{align*}
Thus charging $t_{\hi}=w_{\hi}^0$ satisfies both high-type IR and low-type IC and extracts the high type's entire continuation value. Revenue is therefore $\pi w_{\hi}^0$.
\end{proof}

\begin{proofof}{\bf Theorem \ref{T:secbest}}
Lemma \ref{L:participation} implies that every nonempty incentive-compatible platform either includes both types or only the high type. The platform which excludes both types is strictly dominated by high-only inclusion, which yields positive revenue $\pi w_{\hi}^0$.

Under the no-holdout condition, Lemma \ref{L:noholdout} implies that $\lo\in\mathcal{M}(\lo)$ in every full-inclusion DSE. Proposition \ref{P:nodist} therefore implies that assortative search uniquely maximizes revenue conditional on full inclusion. If only the high type is included, market clearing requires $g_{\hi}=\delta_{\hi}$, so this allocation is also assortative conditional on participation and has zero mismatch. Lemma \ref{L:participation} establishes feasibility of both candidate allocations. Taking the better of their two revenues gives an optimum, and every optimum has $\epsilon=0$.\hfill$\square$
\end{proofof}

\begin{proofof}{\bf Corollary \ref{C:char}}
By Theorem \ref{T:secbest}, only two allocations can be optimal: assortative full inclusion and high-only inclusion. By Lemmas \ref{L:revenue-screening} and \ref{L:participation}, assortative full inclusion yields
\begin{align*}
    (1-\pi)w_{\lo}^0
    +\pi\left[
    w_{\hi}^0-
    \frac{\theta u^0}{1+\theta u^0}
    (f_{\lo\hi}-f_{\lo\lo})
    \right],
\end{align*}
whereas high-only inclusion yields $\pi w_{\hi}^0$. Full inclusion is therefore preferred exactly when
\begin{align*}
    (1-\pi)w_{\lo}^0
    \geq
    \pi\frac{\theta u^0}{1+\theta u^0}
    (f_{\lo\hi}-f_{\lo\lo}).
\end{align*}
Using
\begin{align*}
    w_{\lo}^0=\frac{\theta u^0}{1+2\theta u^0}f_{\lo\lo}
\end{align*}
and cancelling $\theta u^0>0$, this inequality is equivalent to
\begin{align*}
    \left[1+(1+\pi)\theta u^0\right]f_{\lo\lo}
    \geq
    \pi(1+2\theta u^0)f_{\lo\hi}.
\end{align*}
Strict inequality selects assortative full inclusion uniquely; the reverse strict inequality selects high-only inclusion uniquely; equality makes both allocations optimal. The transfer formulas for full inclusion follow from Lemma \ref{L:revenue-screening}, and the high-only transfer follows from Lemma \ref{L:participation}.\hfill$\square$
\end{proofof}

\begin{proofof}{\bf Corollary \ref{C:optimal-efficient}}
Under the stated strict inequality, Corollary \ref{C:char} implies that the unique optimal platform has full inclusion and assortative search. Corollary \ref{C:efficientsearch} says that this is also the unique efficient platform. Hence the unique optimal platform solves \eqref{E:SPP}.\hfill$\square$
\end{proofof}

\begin{proofof}{\bf Proposition \ref{P:paymenttiming}}
Since both own-type meetings are accepted, there are two cases. Either cross-type meetings are also accepted or they are rejected.

Suppose first that every meeting is accepted. As in the baseline model, $u_{\lo}=u_{\hi}=u^0$. Multiplying the two Bellman equations by the corresponding population shares and using market clearing gives
\begin{align*}
    &(1-\pi)w_{\lo}+\pi w_{\hi}\\
    &\qquad=
    \frac{\theta u^0}{1+2\theta u^0}
    \Big[
        (1-\pi)(f_{\lo\lo}-2\kappa_{\lo})
        +\pi(f_{\hi\hi}-2\kappa_{\hi})
        -\epsilon
        (f_{\lo\lo}+f_{\hi\hi}-2f_{\lo\hi})
    \Big].
\end{align*}
The corresponding assortative expression is obtained by setting $\epsilon=0$. Strict supermodularity therefore gives the first inequality, strictly whenever $\epsilon>0$.

Now suppose cross-type meetings are rejected. The balance equations are unchanged by payment timing:
\begin{align*}
    \alpha(1-u_{\lo})
    &=
    \rho(1-a)u_{\lo}^2,\\
    \alpha(1-u_{\hi})
    &=
    \rho(1-b)u_{\hi}^2.
\end{align*}
For $\epsilon>0$,
\begin{align*}
    (1-a)u_{\lo}<u^0,
    \qquad
    (1-b)u_{\hi}<u^0.
\end{align*}
The Bellman equations imply
\begin{align*}
    w_{\lo}
    &=
    \frac{\theta(1-a)u_{\lo}}
    {1+2\theta(1-a)u_{\lo}}
    (f_{\lo\lo}-2\kappa_{\lo}),\\
    w_{\hi}
    &=
    \frac{\theta(1-b)u_{\hi}}
    {1+2\theta(1-b)u_{\hi}}
    (f_{\hi\hi}-2\kappa_{\hi}).
\end{align*}
Own-type acceptance implies that both terms in parentheses are positive. Hence $w_\lo<w_\lo^0$ and $w_\hi<w_\hi^0$ whenever $\epsilon>0$. This proves the first inequality.

It remains to compare informational rents. Write
\begin{align*}
    d
    =
    \tilde w_{\hi,\lo}-w_{\lo},
    \qquad
    d^0
    =
    \frac{\theta u^0}{1+\theta u^0}
    (f_{\lo\hi}-f_{\lo\lo}),
\end{align*}
and define
\begin{align*}
    \Delta_{\lo}
    &=
    f_{\lo\hi}-f_{\lo\lo},\\
    \Delta_{\hi}
    &=
    f_{\hi\hi}-f_{\lo\hi}.
\end{align*}
Strict supermodularity implies $\Delta_\hi>\Delta_\lo$.

First consider the accepting regime. Let
\begin{align*}
    S_{\lo}
    &=
    f_{\lo\lo}
    -2\kappa_{\lo}
    -2w_{\lo}>0,\\
    S_{\hi}
    &=
    f_{\lo\hi}
    -\kappa_{\lo}-\kappa_{\hi}
    -w_{\lo}-w_{\hi}>0.
\end{align*}
Subtracting the truthful low-type Bellman equation from the high type's deviation equation gives
\begin{align*}
    d
    =\theta u^0\Big[
        &(1-a)
        \left(
        [S_{\lo}+\Delta_{\lo}-d]_+
        -S_{\lo}
        \right)+
        a
        \left(
        [S_{\hi}+\Delta_{\hi}-d]_+
        -S_{\hi}
        \right)
    \Big].
\end{align*}
For any $S\geq0$,
\begin{align*}
    [S+x]_+-S\geq x.
\end{align*}
Therefore
\begin{align*}
    d
    \geq
    \theta u^0
    \left[
        (1-a)(\Delta_{\lo}-d)
        +a(\Delta_{\hi}-d)
    \right]
    \geq
    \theta u^0(\Delta_{\lo}-d).
\end{align*}
Hence $d\geq d^0$, with strict inequality whenever $a>0$.

Now consider the selective regime. Define
\begin{align*}
    q=\theta(1-a)u_{\lo},
    \qquad
    p=\theta a u_{\hi}.
\end{align*}
For $\epsilon>0$, both unmatched fractions exceed $u^0$, so $q+p>\theta u^0$. Let
\begin{align*}
    S_{\lo}
    =
    f_{\lo\lo}
    -2\kappa_{\lo}
    -2w_{\lo}>0.
\end{align*}
A high type who chooses the low menu obtains match surplus from a high partner, before accounting for the increase $d$ in his continuation value, equal to
\begin{align*}
    T
    =
    f_{\hi\hi}
    -\kappa_{\lo}-\kappa_{\hi}
    -w_{\lo}-w_{\hi}.
\end{align*}
Own-type acceptance gives
\begin{align*}
    \kappa_{\lo}+w_{\lo}
    <\frac12f_{\lo\lo},
    \qquad
    \kappa_{\hi}+w_{\hi}
    <\frac12f_{\hi\hi}.
\end{align*}
Consequently,
\begin{align*}
    T
    >
    \frac12(f_{\hi\hi}-f_{\lo\lo})
    >
    f_{\lo\hi}-f_{\lo\lo}
    =
    \Delta_{\lo},
\end{align*}
where the second inequality follows from strict supermodularity.

Subtracting the low type's Bellman equation from the deviation equation gives
\begin{align*}
    d
    =
    q\left(
        [S_{\lo}+\Delta_{\lo}-d]_+
        -S_{\lo}
    \right)
    +
    p[T-d]_+.
\end{align*}
The preceding inequalities imply
\begin{align*}
    d
    \geq
    (q+p)(\Delta_{\lo}-d).
\end{align*}
Hence
\begin{align*}
    d
    \geq
    \frac{q+p}{1+q+p}\Delta_{\lo}
    >
    \frac{\theta u^0}{1+\theta u^0}\Delta_{\lo}
    =
    d^0.
\end{align*}
This proves the rent comparison.

Finally, suppose $\widehat f_{xy}$ is strictly increasing. The truthful matching problem is then exactly the baseline matching problem with production function $\widehat f$. Since payment timing preserves strict supermodularity, Lemma \ref{L:regimes} applies to $\widehat f$. Thus $\lo\in\mathcal M(\lo)$ implies $\hi\in\mathcal M(\hi)$, and the preceding argument applies.\hfill$\square$
\end{proofof}

\begin{proofof}{\bf Corollary \ref{C:searchfee}}
Let $U_x$ denote the net continuation value of a type $x$ agent while unmatched. Under a search-only fee $t_x\geq0$, the Bellman equation is
\begin{align*}
    U_x
    =
    -t_x
    +
    \theta
    \sum_{y\in\mathcal M(x)}
    u_y g_x(y)
    \left(
        f_{xy}-U_x-U_y
    \right).
\end{align*}
Thus a meeting between types $x$ and $y$ is accepted if and only if $f_{xy}-U_x-U_y>0$.

I first show that low-low meetings are accepted in every full-inclusion DSE. Suppose to the contrary that they are rejected. The characterization of matching regimes then implies $\mathcal{M}(\lo)=\{\hi\}$ and $\mathcal{M}(\hi)=\{\lo,\hi\}$. The low type's Bellman equation is
\begin{align*}
    U_{\lo}
    =
    -t_{\lo}
    +
    \theta a u_{\hi}
    \left(
        f_{\lo\hi}-U_{\lo}-U_{\hi}
    \right).
\end{align*}
Since $t_{\lo}\geq0$ and individual rationality implies $U_{\hi}\geq0$,
\begin{align*}
    U_{\lo}
    \leq
    \theta a u_{\hi}
    \left(
        f_{\lo\hi}-U_{\lo}
    \right).
\end{align*}

The balance equations are unchanged by the payment rule. The argument in the proof of Lemma \ref{L:noholdout} therefore gives $a u_{\hi}\leq u^0$. Hence,
\begin{align*}
    U_{\lo}
    \leq
    \theta u^0
    \left(
        f_{\lo\hi}-U_{\lo}
    \right),
\end{align*}
and therefore
\begin{align*}
    U_{\lo}
    \leq
    \frac{\theta u^0}{1+\theta u^0}
    f_{\lo\hi}.
\end{align*}
Low-low rejection requires $f_{\lo\lo}\leq2U_{\lo}$, so
\begin{align*}
    f_{\lo\lo}
    \leq
    \frac{2\theta u^0}{1+\theta u^0}
    f_{\lo\hi},
\end{align*}
contradicting the no-holdout condition. Thus $\lo\in\mathcal{M}(\lo)$ in every full-inclusion DSE. Proposition \ref{P:paymenttiming} now applies.\hfill$\square$
\end{proofof}

\begin{proofof}{\bf Proposition \ref{P:searchintensity}}
Fix a platform associated with a selective DSE and define
\begin{align*}
    \widehat\rho_x=\rho_xg_x(x).
\end{align*}
Since cross-type meetings are rejected, the truthful Bellman and balance equations depend on $(g_x,\rho_x)$ only through $\widehat\rho_x$. Replacing the original platform by
\begin{align*}
    \{(\delta_\lo,\widehat\rho_\lo,t_\lo),
      (\delta_\hi,\widehat\rho_\hi,t_\hi)\}
\end{align*}
therefore leaves truthful reservation wages and unmatched fractions unchanged.

The replacement can only reduce deviation values. For example, a high type choosing the low menu receives low-type opportunities at exactly the same rate as before, but no longer receives any high-type opportunities generated by $g_\lo(\hi)$. Hence
\begin{align*}
    \tilde w_{\hi,\lo}^{\,\prime}
    \leq
    \tilde w_{\hi,\lo},
\end{align*}
and similarly for the upward deviation. Thus assortative search weakly relaxes incentive compatibility while leaving truthful values unchanged.

It remains to choose search intensities. Define $\lambda_x=\rho_xu_x$. Under assortative search,
\begin{align*}
    u_x
    =
    \frac{\alpha}{\alpha+\lambda_x},
    \qquad
    w_x
    =
    \frac{\lambda_x}
         {2(r+\alpha+\lambda_x)}
    f_{xx},
\end{align*}
and the high type's downward information rent is
\begin{align*}
    \tilde w_{\hi,\lo}-w_\lo
    =
    \frac{\lambda_\lo}
         {2(r+\alpha)+\lambda_\lo}
    (f_{\lo\hi}-f_{\lo\lo}).
\end{align*}
Strict supermodularity implies that, whenever $\lambda_\hi\geq\lambda_\lo$, the upward IC constraint is slack. Revenue is therefore
\begin{align*}
    &(1-\pi)
    \frac{\lambda_\lo f_{\lo\lo}}
         {2(r+\alpha+\lambda_\lo)}
    +
    \pi
    \frac{\lambda_\hi f_{\hi\hi}}
         {2(r+\alpha+\lambda_\hi)}
    -
    \pi
    \frac{\lambda_\lo(f_{\lo\hi}-f_{\lo\lo})}
         {2(r+\alpha)+\lambda_\lo}.
\end{align*}
The high-type term is strictly increasing in $\lambda_\hi$, while the informational rent does not depend on $\lambda_\hi$. Increasing $\lambda_\hi$ also makes cross-type matching less attractive. Hence $\lambda_\hi=\bar\rho u^0$, leaving the stated one-dimensional problem for $\lambda_\lo$.\hfill$\square$
\end{proofof}

\begin{proofof}{\bf Corollary \ref{C:selectiveintensity}}
Any nonselective full-inclusion DSE must involve cross-type acceptance. Hence $w_\lo+w_\hi<f_{\lo\hi}$.  Individual rationality implies $t_x\leq w_x$, so its revenue is bounded by
\begin{align*}
    (1-\pi)w_\lo+\pi w_\hi
    <
    \max\{\pi,1-\pi\}f_{\lo\hi}.
\end{align*}
The first condition therefore rules out every such platform. The second compares the optimal selective platform with high-only inclusion.\hfill$\square$
\end{proofof}

\end{document}